\documentclass[twocolumn,times]{aastex631}
\usepackage{amssymb}
\received{\today}
\revised{}
\accepted{}
\submitjournal{ApJ}

\usepackage{color,comment}
\usepackage[normalem]{ulem}
\usepackage{newtxtext,newtxmath}
\usepackage{amsmath} 
\usepackage{ctable}

\newcommand{\angstrom}{\textup{\AA}}

\newcommand{\qpah}[1]{$q_{\rm PAH}$ }
\newcommand{\lpah}[1]{$L_{\rm PAH}$ }
\newcommand{\lfir}[1]{$L_{\rm FIR}$ }

\shorttitle{sharing all the beta you need about $\beta$}
\shortauthors{Narayanan et al.}

\begin{document}

\title[]{The Ultraviolet Slopes of Early Universe Galaxies: \\ The Impact of Bursty Star Formation, Dust, and Nebular Continuum Emission}

\correspondingauthor{Desika Narayanan}
\email{desika.narayanan@ufl.edu}

\author[0000-0002-7064-4309]{Desika Narayanan}
\affil{Department of Astronomy, University of Florida, 211 Bryant Space Sciences Center, Gainesville, FL 32611 USA}
\affil{Cosmic Dawn Center at the Niels Bohr Institute, University of Copenhagen and DTU-Space, Technical University of Denmark}
\author{Daniel P. Stark}
\affil{Steward Observatory, University of Arizona, 933 N Cherry Ave, Tucson, AZ, 85721, USA}

\author[0000-0001-8519-1130]{Steven L. Finkelstein}
\affil{Department of Astronomy, The University of Texas at Austin, Austin, TX 78712, USA}

\author[0000-0002-5653-0786]{Paul Torrey}
\affil{Department of Astronomy, University of Virginia, 530 McCormick Road, Charlottesville, VA 22903, USA}

\author[0000-0001-8015-2298]{Qi Li}
\affil{Max Planck Institute for Astrophysics, Garching bei Munchen, Germany}

\author[0000-0002-3736-476X]{Fergus Cullen}
\affil{Institute for Astronomy, University of Edinburgh, Royal Observatory, Edinburgh, EH9 3HJ, UK}

\author [0000-0001-8426-1141]{Micheal W. Topping}
\affil{Steward Observatory, University of Arizona, 933 N Cherry Ave, Tucson, AZ, 85721, USA}

\author[0000-0003-3816-7028]{Federico Marinacci}
\affil{Department of Physics and Astronomy "Augusto Righi", University of Bologna, via Gobetti 93/2, 40129, Bologna, Italy}
\affiliation{INAF, Astrophysics and Space Science Observatory Bologna, Via P. Gobetti 93/3, 40129 Bologna, Italy}

\author[0000-0002-3790-720X]{Laura V. Sales}
\affiliation{Department of Physics and Astronomy, University of California, Riverside, CA, 92521, USA}

\author[0000-0002-6196-823X]{Xuejian Shen}
\affiliation{Department of Physics, Kavli Institute for Astrophysics and Space Research, Massachusetts Institute of Technology, Cambridge, MA 02139, USA}

\author[0000-0001-8593-7692]{Mark Vogelsberger}
\affiliation{Department of Physics, Kavli Institute for Astrophysics and Space Research, Massachusetts Institute of Technology, Cambridge, MA 02139, USA}




\begin{abstract}
 
JWST has enabled the detection of the ultraviolet (UV) continuum of galaxies at $z>10$, evidencing a population of extremely blue,
potentially dust-free galaxies.  Interpreting the UV spectra of galaxies as they redden is
complicated by the well-known degeneracy between stellar ages, dust
  reddening, and nebular continuum.
  The main goal of this paper is to develop a theoretical model for
  the relationship between galaxy UV slopes ($\beta$), bursty star
  formation histories, dust evolution, and the contribution
  from nebular regions.  We accomplish this via cosmological zoom-in
 simulations, and in specific, build a layered model where we simulate the UV slopes of galaxies with increasingly complex physics.  In particular, we study the UV slopes of (i) unattenuated intrinsic stellar populations in high-redshift galaxies, (ii) reddened populations using a new on-the-fly evolving dust model, and (iii) stellar populations that include both dust and the contribution of nebular continuum.  
Our main results follow.  Unattenuated stellar populations with no nebular emission exhibit a
diverse range of intrinsic UV slopes, with values ranging from $\beta_0
\approx  -3\rightarrow-2.2$ due to long delays between bursts.  This is manifested by
an inverse correlation between the intrinsic UV slope and sSFR for
early galaxies such that higher sSFR corresponds to bluer UV slopes. When including dust, our model galaxies demonstrate a rapid rise in
dust obscuration between $z\approx 8-10$.  This increase in dust mass is due to high
grain-grain shattering rates, and enhanced growth per unit dust mass
in very small grains, resulting in UV-detected galaxies at $z \sim 12$
descending into ALMA-detectable galaxies by $z \sim 6$.  The rapid rise in dust content at $z\approx8-10$ leads to a systematic reddening of the UV slopes during this redshift range. 
 The
inclusion of nebular continuum reddens the UV slope by a median factor
$\Delta \beta_{\rm neb} \approx 0.2-0.4$.  However, when including
nebular continuum, our highest redshift galaxies ($z\approx12$) are
insufficiently blue compared to observations; this may imply an
evolving escape fraction from HII regions with redshift.

\end{abstract}




\section{Introduction}
\label{section:introduction}

 The rest-frame ultraviolet (UV) continuum power-law slope of galaxies
 has long been used as a diagnostic for the physical conditions of
 high-redshift galaxies \citep[see the reviews by][and
   references therein]{dunlop13a,finkelstein16a,stark16a,dayal18a,robertson22a}.
 Formally defined by $f_\lambda \propto \lambda^\beta$ \citep{calzetti94a}, the slope
 $\beta$ is driven by a combination of the intrinsic ultraviolet
 continuum of unreddened stellar populations
 \citep[][]{reddy18a,calabro21a}, potential contributions by nebular
 continuum
 \citep{byler17a,cullen17a,chisholm22a,topping22a,topping24a}, and
 reddening by dust extinction and attenuation \citep[e.g.][and
   references therein]{salim20a}.

For the vast majority of dust-enriched, star-forming galaxies, any
deviation in $\beta$ from the intrinsic stellar UV continuum slope
(which we refer to hereafter as $\beta_0$) is dominated by dust
obscuration.  Seminal work by \citet{calzetti97a} and
\citet{meurer99a} found that local starburst galaxies lie on a
well-defined plane between the infrared excess (IRX: ratio of infrared
luminosity to UV luminosity) and $\beta$, which can be translated to a
family of dust-attenuation relations
\citep[e.g.][]{siana09a,popping17a,narayanan18a,shen20a,liang21a}.  As a
result, a number of studies have employed optical and near-infrared
photometry to probe the rest-frame UV slopes of galaxies from
$z\approx2-9$ in order to investigate the potential evolution of dust
attenuation properties of galaxies during the first $\sim 3-4$ billion
years of the Universe's history.  The general sense from these surveys
is that galaxies toward the Epoch of Reionization (EoR; $z>6$) are
systematically bluer in their UV slopes than galaxies at $z
\sim 2$, with the inference that dust attenuation becomes less
impactful at earlier times
\citep[e.g.][]{stanway05a,bouwens10a,finkelstein10a,mclure11a,bouwens12a,dunlop12a,finkelstein12a,bouwens14b}. At
a fixed redshift, some studies additionally find bluer UV slopes with decreasing UV luminosity \citep{bouwens12a,cullen23a,tacchella22a},
which is natural in a scenario where more massive galaxies\footnote{Indeed, \citet{morales24b} have shown an even stronger dependence of $\beta$ on galaxy mass.} have
elevated dust content
\citep[e.g.][]{bethermin15a,dacunha15a,donevski20a}.  The inferred
rise in dust content from rest-frame UV continuum observations
from $z=8\rightarrow2$ is complemented by depletion measurements
\citep{peroux20a,peroux23a}, extinction measurements
\citep{menard10a,menard12a}, SED fits to galaxies
\citep{dunne11a,beeston18a,driver18a,pozzi20a}, and theoretical models
\citep{aoyama18a,li18a}, all of which point towards a rise in the cosmic dust
density from early times toward cosmic noon ($z \sim 2$) \citep{peroux20a}.

While some limited samples of UV detections of even higher redshift
($z>10$) galaxies have been studied in the pre-JWST era
\citep[e.g.][]{ellis13a,oesch16a,wilkins16a,tacchella22a}, the sheer
number of rest-frame UV detections at $z>10$ has exploded since the
successful launch of JWST/NIRCam
\citep[e.g.,][]{naidu22a,curtislake23a,donnan23a,donnan23b,finkelstein23b,finkelstein23a,harikane23a,robertson23a,casey24a,ciesla24a,roberts-borsani24a}.
This has enabled the investigation of the redshift evolution of the UV
slope $\beta$ to increasingly early times, where the impact of dust obscuration may become relatively minimal.  To wit, recent observations
by \citet{cullen24a} \citet{topping24a}, and \citet{morales24a} found average UV slopes of value
$\langle \beta \rangle \approx -2.5$ to -$2.8$ at redshifts $z>10$, which is consistent with
their expectations of dust-free stellar populations within the
first $500$ Myr given the specifics of their population synthesis modeling.

At the same time, Atacama Large Millimetre Array (ALMA) observations
of galaxies at $z \sim 6$ have demonstrated significant dust
reservoirs by $z=6$ \citep[e.g.][]{pozzi21a,bouwens22a,inami22a}, with
dust to stellar mass ratios $\sim 10^{-3}$, already comparable to
lower-redshift systems \citep{donevski20a}.  If the JWST UV-detected
$z>10$ galaxies are related to the ALMA REBELS $z\approx 6$ galaxies
(and indeed, in this paper, we demonstrate that such a connection may
be credible\footnote{A very rough back-of-the-envelope calculation
additionally suggests that a causal connection between the JWST
UV-detected galaxies at $z>10$ and ALMA-detected galaxies at $z\sim6$
may be reasonable.  The lower bound of galaxies with
dust-to-stellar ratios of $10^{-3}$ at $z \sim 6$ have stellar masses
$\sim 10^9$ M$_\odot$ \citep[][see also Figure~\ref{figure:rebels} in this paper as well]{topping22a}.  At the same time, some
studies have inferred stellar masses $\sim 10^8 $M$_\odot$ for $z \sim
9-10$ UV selected galaxies \citep[][acknowledging, of course, the challenges in inferring stellar masses in potentially bursty systems at such high redshifts; \citealt{tacchella22a,whitler23a,ciesla24a,narayanan24a}]{morales24a}.  This implies a stellar growth of $\sim 10^9 $M$_\odot$ in $\sim 500$
Myr, which is achievable with a SFR $\sim 2 $M$_\odot$ yr$^{-1}$. Main
sequence star formation for these stellar masses is already SFR$\sim 1-10
$M$_\odot$ yr$^{-1}$ at these redshifts \citep{speagle14a,lower23a}.}),
then this would imply a sudden onset of dust obscuration in the
relatively short time between $z=10\rightarrow 6$.  How galaxies go
from inferred dust-free stellar populations to significantly dust-enriched in only $\sim 0.5$ Gyr represents a theoretical challenge
\citep[e.g.][]{ferrara23a,ziparo23a}.  The observed inference of the
rise of dust obscuration is further complicated by the potential
impact of intrinsically red UV slopes due to aged stellar populations
\citep[e.g.][]{topping22a,cullen24a}, as well as nebular continuum
emission from free-free and free-bound processes \citep{byler17a,katz24a}.

The main goal of this paper is to develop a theoretical model for the
relationship between galaxy UV slopes, dust growth, bursty star
formation histories, emission from nebular regions, and cosmological
galaxy growth in the early Universe.  We accomplish this via a
combination of cosmological zoom-in simulations (in order to generate
a set of model galaxies to study), a new model for the formation,
growth, and evolution of multi-species and multi-sized dust, and dust
radiative transfer calculations (in order to generate mock
observations).  In \S~\ref{section:methods}, we describe these methods
in detail.  In \S~\ref{section:physical_properties}, we take the
reader on a tour of the physical properties of our model galaxies to
help place them in a broader context.  We follow this in
\S~\ref{section:unreddened} with a study of the UV properties of
intrinsic, unreddened stellar populations: this is important as aging
stellar populations redden, resulting in degenerate solutions with the
impact of dust and nebular continuum.  We follow this in
\S~\ref{section:reddened} with an exploration of the onset of dust
obscuration in the early Universe.  In \S~\ref{section:nebular}, we
investigate how the emission from nebular regions impacts the UV
slopes of our model galaxies.  In \S~\ref{section:discussion} we
provide discussion, and in \S~\ref{section:summary} we summarize.

Finally, as discussed previously in this section, there are a range of
potential physical effects on $\beta$ that we will investigate in this
paper.  In an attempt to avoid confusion, we summarize our working
definitions of the various forms of $\beta$ in
Table~\ref{table:beta_def}.  For all mock SEDs, we fit our SEDs at the
center of the \citet{calzetti94a} wavelength windows.

\begin{table*}
        \centering
        \caption{$\beta$ Definitions Used in this Paper}
        \label{table:beta_def}
        \begin{tabular}{l|r}
                \hline
                $\beta$ & Generic UV slope of a galaxy\\
                $\beta_0$ & UV slope of intrinsic stellar populations (i.e. with no dust or nebular continuum) \\
                $\beta_{\rm dust}$& UV slope of galaxy that includes dust, but not nebular continuum\\
                $\beta_{\rm neb}$& UV slope of a galaxy that includes dust as well as nebular continuum \\
                $\Delta \beta_{\rm dust}$ & $\beta_{\rm dust}$-$\beta_0$ (isolating the impact of dust)\\
                $\Delta \beta_{\rm neb}$ & $\beta_{\rm neb}$-$\beta_{\rm dust}$ (isolating the impact of nebular continuum)\\
                \hline
                
        \end{tabular}
\end{table*}

\section{Simulation Strategy and Methodology}
\label{section:methods}

\subsection{Galaxy Formation Models}
We run a suite of cosmological zoom-in simulations of massive galaxies
down to $z=6$ using the {\sc smuggle} galaxy formation model within
the moving mesh {\sc arepo} hydrodynamic and gravity code
\citep{springel10a,weinberger20a}.  The {\sc smuggle} physics model is
described fully in \citet{marinacci19a}, and we refer the reader to
this paper for an in-depth discussion of the model; we summarize the
salient points here.

Primordial cooling of the gas occurs via two-body collisional
processes, free-free emission, recombination, and Compton cooling off
of cosmic microwave background photons \citep{katz96a}.  Once enriched
with metals, gas can additionally undergo metal line cooling with
rates computed as a function of temperature and density via {\sc
  cloudy} photoionization calculations \citep{ferland13a} as described
in \citet{vogelsberger13a}.  Low-temperature gas cools via a range of
processes, including metal-line cooling, fine-structure emission, and
molecular cooling via fits to the \citet{hopkins18a} {\sc cloudy}
cooling tables \citep{marinacci19a}.  Self-shielding occurs in gas
with densities $n>10^{-3} $cm$^{-3}$ via the \citet{rahmati13a}
redshift-dependent parameterization.  Cosmic ray heating follows the
density-dependent prescription of \citet{guo08a}, and photoelectric
heating follows the \citet{wolfire03a} density, metallicity, and
temperature-dependent rates.

Star formation occurs in gravitationally bound molecular gas
\citep{hopkins13a} above a density threshold $n_{\rm thresh}=100$
cm$^{-3}$, following a volumetric \citet{kennicutt98a} relation
$\dot{M_*} = \epsilon M_{\rm gas}/t_{\rm ff}$, where $\dot{M_*}$ is
the star formation rate (SFR), $M_{\rm gas}$ is the gas mass, and
$t_{\rm ff}$ is the free fall time.  Following \citet{marinacci19a},
we set the star formation efficiency $\epsilon=0.01$
\citep{krumholz07b,krumholz10a}.  The molecular fraction of the
interstellar medium (ISM) is computed via the \citet{krumholz08a}
prescription linking the H$_2$ fraction to the local gas surface
density and metallicity.

Once formed, stars return energy to the nearby ISM via a range of
feedback processes.  We compute the fraction of supernovae (SNe) from
each formed star particle assuming a \citet{chabrier01a} 
initial mass function, with an additional delay time distribution for
deriving the number of Type Ia supernovae events
\citep{vogelsberger13a, torrey14a}.  The details for the SNe-driven stellar mass
loss rates, as well as the coupling of the energy and momentum from
SNe ejecta to the ISM are provided in \citet{marinacci19a} and
\citet{zhang24a}.  Similarly, feedback is included via radiation from
young stars via photoionization, radiation pressure, and OB and AGB
stellar winds.  Taken together, these feedback processes regulate the
star formation rates in our model zoom-in galaxies, as well as the
physical conditions in the ISM.  One major consequence of the default
{\sc smuggle} galaxy formation framework that we adopt is bursty star
formation histories (SFHs), which may be reasonable for early Universe
galaxies per observational constraints
\citep{ciesla23a,dressler23a,endsley23a,looser23a,shen23a}.  This said, it is
worth noting that the magnitude of this burstiness may be sensitive to
the details of unresolved SNe feedback physics \citep{zhang24a}.

In addition to injecting feedback, aging stellar populations return mass and metals to the ISM. 
The enrichment of the ISM is tracked for 9 individual elements, as well as metal content as a whole.
Metals that are present in the ISM are advected as passive tracers with the fluid flow.
This approach allows for self-consistent co-evolution of galaxies and their metal content including, e.g., the emergence of a time and mass dependent mass-metallicity relation and internal metallicity gradients. 
The locally resolved free abundance of metals, in turn, can impact/modulate the accretion rates of metals onto dust grains.

\begin{table*}
        \centering
        \caption{Physical properties of model zoom-in simulations considered in this paper.}
        \label{table:zooms}
        \begin{tabular}{l|lllll}
                \hline
                Galaxy Name & M$_{\rm DM}$ & M$_{\rm DM}$ & M$_*$ & M$_{\rm gas}$ & M$_{\rm dust}$ \\
                & M$_\odot$; $z=0$ &  M$_\odot$; $z=6$ &  M$_\odot$; $z=6$ &  M$_\odot$; $z=6$ &  M$_\odot$; $z=6$\\
                \hline
                h10 &$6.2 \times 10^{13}$&$3.2 \times 10^{11}$& $3.7 \times 10^9$&$2.0 \times 10^{10}$&$3.5 \times 10^7$\\
                h15 &$5.3 \times 10^{13}$&$4.1 \times 10^{11}$&$1.8 \times 10^{10}$&$1.5 \times 10^{10}$&$1.8 \times 10^8$\\
                h17 & $4.1 \times 10^{13}$ & $3.3 \times 10^{11}$& $7.2 \times 10^9$& $1.7 \times 10^{10}$& $6.0 \times 10^7$\\
                h20 & $3.5 \times 10^{13}$ &$2.5 \times 10^{11}$&$3.15 \times 10^9$&$1.4 \times 10^{10}$&$2.8 \times 10^7$\\
                h25 &$2.6 \times 10^{13}$&$1.8 \times 10^{11}$&$3.9 \times 10^9$& $1.1 \times 10^{10}$&$3.8 \times 10^7$\\
                \hline
                
        \end{tabular}
\end{table*}

\subsection{Dust Physics}
In order to investigate the buildup of dust reservoirs in early
galaxies, and their obscuring impact on stellar populations, we
require a model for dust formation, growth, and destruction in
galaxies.  In particular, to properly model the local extinction of UV
photons from individual star particles in our simulations (as well as
the global attenuation), we need to know the local dust mass, grain
size distribution, and chemical composition.  To do this, we have
implemented the newly developed dust model of \citet{li21a} and
\citet{narayanan23a} on the fly in cosmological zoom-in simulations, which build on the algorithms developed by \citet{mckinnon16a,mckinnon17a,mckinnon18a}.
We refer the reader to \citet{narayanan23a} for details of the
coupling of our dust model with the {\sc arepo smuggle} galaxy
formation model, and summarize the salient points here.

Dust is produced through the condensation of metals that are ejected
from evolved stars.  The dust yields are taken from
\citet{schneider14a} for AGB stars, and \citet{nozawa10a} for SNe
dust production.  This dust is initialized with a lognormal size
distribution, though dust particles lose their memory of these initial
conditions rapidly owing to coagulation and shattering processes (more
on this below).  Each dust particle is discretized into $16$
logarithmic size bins in the range $4 \times 10^{-4} \leq (a/\mu{\rm m}) \leq 1$.  The mass
of dust produced by evolved stars follows the methodology of
\citet{dwek98a}, with condensation efficiencies taken from
\citet{ferrarotti06a} for AGB stars, and \citet{bianchi07a} for SNe.
We assume a silicate-carbonaceous chemical breakdown of our dust
particles, where the total carbon mass in a dust particle corresponds
to the carbonaceous dust mass, and the remainder to silicate.

Dust grains evolve from their initialized state by growing mass and
size via metal accretion and coagulation, and being destroyed via
shattering events and shocks in star-forming regions.  The growth timescale is 
inversely proportional to the metal density and the square root of the
temperature:
 \begin{equation}
   \label{equation:growth}
\tau_{\rm accr} = \tau_{\rm ref}\left(\frac{a}{0.1 \mu {\rm m}}\right)\left(\frac{1000 \ {\rm cm}^{-3} \times  m_{\rm H} \times Z_\odot}{\rho_Z}\right)\left(\frac{10{\rm K}}{T_g}\right)^{\frac{1}{2}}\left(\frac{0.3}{S}\right)
 \end{equation}
where $\rho_Z$ is the metal density, $T_g$ is the gas temperature, and
$S$ is the sticking coefficient. We adopt ($\tau_{\rm
  ref},Z_{\odot,{\rm SI}}$) = $\left(0.224 \ {\rm Gyr}, 7 \times
10^{-3}\right)$ for silicates \citep[assuming a composition of
  MgFeSiO$_4$ for silicates;][]{weingartner01a}, and ($\tau_{\rm
  ref},Z_{\odot,{\rm C}}$) = $\left(0.175 \ {\rm Gyr}, 2.4 \times
10^{-2}\right)$ for carbonaceous grains.  We additionally adopt
temperature dependent sticking coefficient following
\citet{zhukovska16a} which drops at higher temperatures. This has the effect of reducing the growth rates in
the warm and dense ISM that is heated by stellar feedback \citep{choban24b}.  Following
\citet{choban22a}, we assume that the growth timescale is limited by the least
abundant element required by the grain species.  Building upon the
framework that we introduced in \citet{narayanan23a}, we have further
updated our dust model to also include size-dependent growth rates
following \citet{weingartner99a}: metals are preferentially depleted
onto very small dust grains owing to enhanced Coulomb potentials.
Functionally, we employ the ISM-phase dependent enhancement vs grain
size curves from \citet{weingartner99a}, with a log-linear
interpolation between the temperature-dependent growth enhancement
rates.  Finally, as will be described in a forthcoming paper
(Narayanan et al., in prep), our cosmological simulations do not
resolve the densities comparable to dense clumps in giant clouds,
where dust growth may be most efficient.  As a result, we assume that
gas is clumped on scales below our resolution limit.  In practice, we
assume that $50\%$ of the mass of a cell has a clumping factor $C=10$,
which impacts all density-dependent rates in our dust model.

We consider the impact of grain-grain collisions on the size
distribution of dust grains.  
There are two important effects: dust shattering in high-speed
collisions, which transforms large grains into smaller grains, and
dust coagulation, which is the vice-versa.  We follow
\citet{mckinnon18a} and \citet{li21a} in modeling the transformation
of grain sizes via collisional encounters, and refer to
\citet{narayanan23a} for the relevant species-dependent equations.  We
employ the default threshold velocities for collisional encounters
from \citet{narayanan23a}, noting that these have resulted in both
model extinction laws in Milky Way analogs comparable to the
\citet{cardelli89a} observational constraints \citep{li21a}, as well
as polycyclic aromatic hydrocarbon (PAH) mass fractions comparable to
constraints from the Local Group \citep{smith07a,narayanan23a}.

Dust can be destroyed via thermal sputtering in the ISM, and via SNe
shocks in star-forming regions.  The former process is
driven by erosion by hot electrons, where the sputtering timescale is
adopted from the analytic approximation of \citet{tsai95a}, and is
linearly proportional to the grain radius, and inversely proportional
to the gas density and temperature.  SNe shocks destroy grains via a
sputtering process as well.  Here, we model the evolution of the grain
size distribution near SNe following the models of dust destruction in
blastwaves by \citet{nozawa06a} and \citet{asano13a} \citep[see ][for
  the relevant equations]{narayanan23a}.

\subsection{Zoom-In Technique}
With the aforementioned {\sc smuggle}+dust physics enabled in the {\sc
  arepo} moving mesh hydrodynamics and gravity solver, we proceed to
simulate the evolution of $5$ galaxies using the well-established
cosmological zoom-in technique \citep[see][for a review of this technique]{vogelsberger20a}. We first run a $\left(50/h\right)^3$
cMpc$^3$ dark matter only simulation with $256^3$ dark matter particles
initialized at $z=99$ down to $z=0$.  The initial conditions are computed with {\sc music} \citep{hahn11a}.  We then select halos of interest
in the $z=0$ snapshot, and construct an ellipsoidal mask around all
particles within $2.5 \times$ the radius of the maximum distance
dark-matter particle in the halo with the {\sc caesar} galaxy analysis
package \citep{thompson14a}.  The region encapsulated by this
ellipsoidal mask is the Lagrangian high-resolution region that will be
re-simulated at higher resolution, and with baryon physics included.
We split the particles within the mask until the effective particle
count in the zoom-in region is $2048^3$ for a baryon (star and gas) mass resolution
in the high-resolution region of $3 \times 10^5 M_\odot$.   Dust particle masses are a factor $\sim 100$ lower mass on average.  These resolution limits are set based on convergence studies for the main physical and mock observable properties that we model here.  We then
re-simulate these halos from the initial conditions ($z=99$) down to
$z=6$.  We summarize the main physical properties of our simulated
galaxies in \S~\ref{section:physical_properties} and
Table~\ref{table:zooms}.

\subsection{Mock Observations}
We generate mock UV SEDs from our model galaxies with the publicly
available {\sc powderday} dust radiative transfer package
\citep{narayanan21a}.  Functionally, {\sc powderday} wraps {\sc yt,
  fsps} and {\sc hyperion} for grid generation, stellar population
synthesis, and Monte Carlo radiative transfer, respectively
\citep{turk11a,conroy10a,robitaille11a}.  For each galaxy snapshot, we
compute the stellar SED for every star particle based on its age and
metallicity with {\sc fsps}. We
assume {\sc mist} stellar isochrones, and a stellar IMF consistent
with the hydrodynamic simulations.

The stellar light is emitted isotropically, and is absorbed,
scattered, and re-emitted by dust in the galaxy.  Here, the radiative
transfer occurs on a Voronoi mesh built around the dust particles with
the active dust model. Informed by the grain size distribution in
every cell, we compute the local extinction law on a cell-by-cell
basis following \citet{li21a}.  In detail, the wavelength-dependent
optical depth seen by every stellar photon can be computed in terms of an extinction efficiency:
\begin{equation}
\tau\left(a,\lambda\right) = \int_{\rm LOS} \pi a^2 Q_{\rm ext}\left(a,\lambda\right) n_d\left(a\right)\mathrm{d}a \ \mathrm{d}s
\end{equation}
where $Q_{\rm ext}$ is the extinction efficiency, d$s$ is the path length of the ray, and $n_{\rm
  d}\left(a\right)$ is the number density of grains with sizes
$\left[a,a+\mathrm{da}\right]$. We assume extinction efficiencies from
\citet{draine84a} and \citet{laor93a} for silicate and carbonaceous
grains, respectively.  

The direction and frequency of photons are randomly drawn from uniform distributions, and the
photons are propagated through the dusty ISM until they either escape
the galaxy, or reach a randomly drawn optical depth drawn from an
exponential distribution.  This process is iterated until the
equilibrium dust temperature has converged (in practice, we iterate until the energy absorbed
by 99\% of the cells has changed by less than 1\%.)  At this point, the aggregate
SED from the model galaxy is computed via ray tracing.   When reporting $\beta$ values, we report the viewing-angle averaged values, noting that there may be some intrinsic dispersion owing to sightline as the galaxies become dustier \citep[e.g.][]{lovell22a,cochrane24a}.

\subsection{Nebular Emission Model}
For a subset of models, we investigate the impact of nebular continuum
emission from star-forming regions.  To do this, we implement the
subresolution model of \citet{garg22a,garg23a}, and refer the reader
to these works for details.  We summarize the important points here.

For all star particles with ages $T_{\rm age} < 10$ Myr, we compute
the shape of the stellar SED from {\sc fsps}, using the stellar age
and metallicity as before.  We assume this SED is incident on a
spherical {\sc HII} region, and use {\sc cloudy v17.00}
\citep{ferland17a} to compute the nebular emission properties. 
We fix
the inner radius of the {\sc HII} region as $R_{\rm inner} = 10^{17}$
cm, following the motivation presented in \citet{garg23a}.  The
hydrogen density\footnote{Some observations have inferred higher nebular densities , during strong bursts at the redshifts of interest here \citep[e.g.][]{topping24a}.   However, as we will discuss in \S~\ref{section:discussion}, this is unlikely to impact the main outcomes of this paper.} of the nebular region is assumed $n_{\rm HII} = 10^2$
cm$^{-2}$, and we model the region as dust-free (in order to focus on
the impact of diffuse dust on $\beta$ in this paper), and the escape
fraction is assumed $f_{\rm esc} = 0$.  Using this setup, we compute
the emergent nebular spectrum from the {\sc HII} region, and then
replace the stellar SED with this SED for propagation through the
dusty ISM with our {\sc powderday} calculations.

\section{The Physical and Observable Properties our Model Early Universe Galaxies}
\label{section:physical_properties}
\begin{figure}
  \includegraphics[scale=0.6]{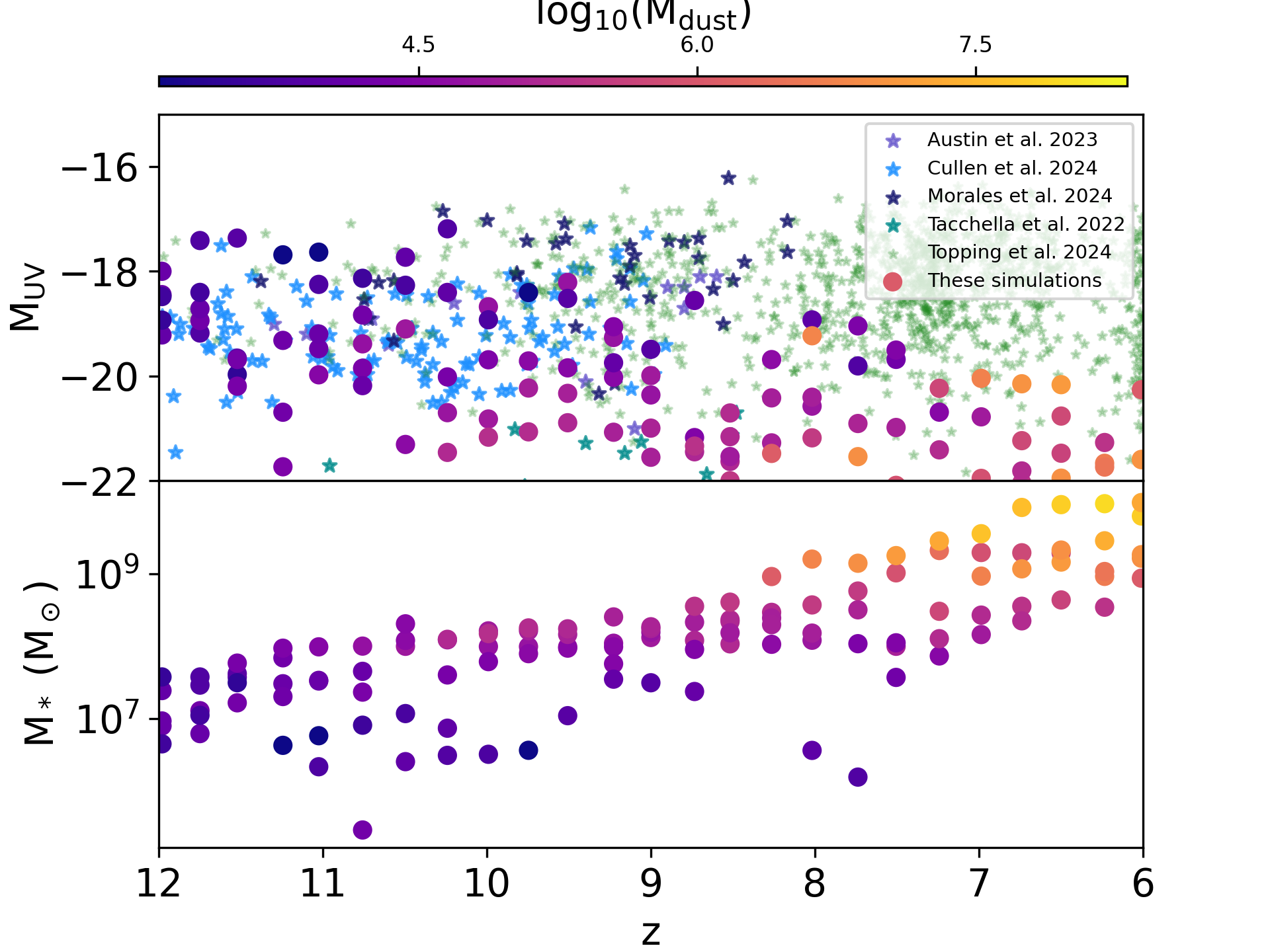}
  \caption{{\bf Observable and physical evolution of model galaxies
      with cosmic time.}  {\it Top:} $M_{\rm UV}$ vs $z$ for model
    galaxies (circles), color-coded by their total dust mass.  In
    comparison are the observational results from
    \citet{austin23a,cullen24a,tacchella22a} and \citet{topping24a}.
    The model galaxy UV luminosities are similar to moderate-area JWST surveys \citep[e.g. CEERS;][]{finkelstein23b}, though brighter than deep fields \citep[e.g. NGDEEP;][]{bagley24a}.   In general, our models are comparable to those from the aforementioned observational samples at $z>10$. This suggests that our model galaxies may serve as reasonable
    analogs for those detected at $z>10$ with HST and JWST, which is
    the primary galaxy sample of interest for this paper.  {\it
      Bottom:} Stellar mass evolution as a function of
    redshift, color-coded by dust mass.  \label{figure:muv_mass_z}}
\end{figure}

\begin{figure}
  \includegraphics[scale=0.6]{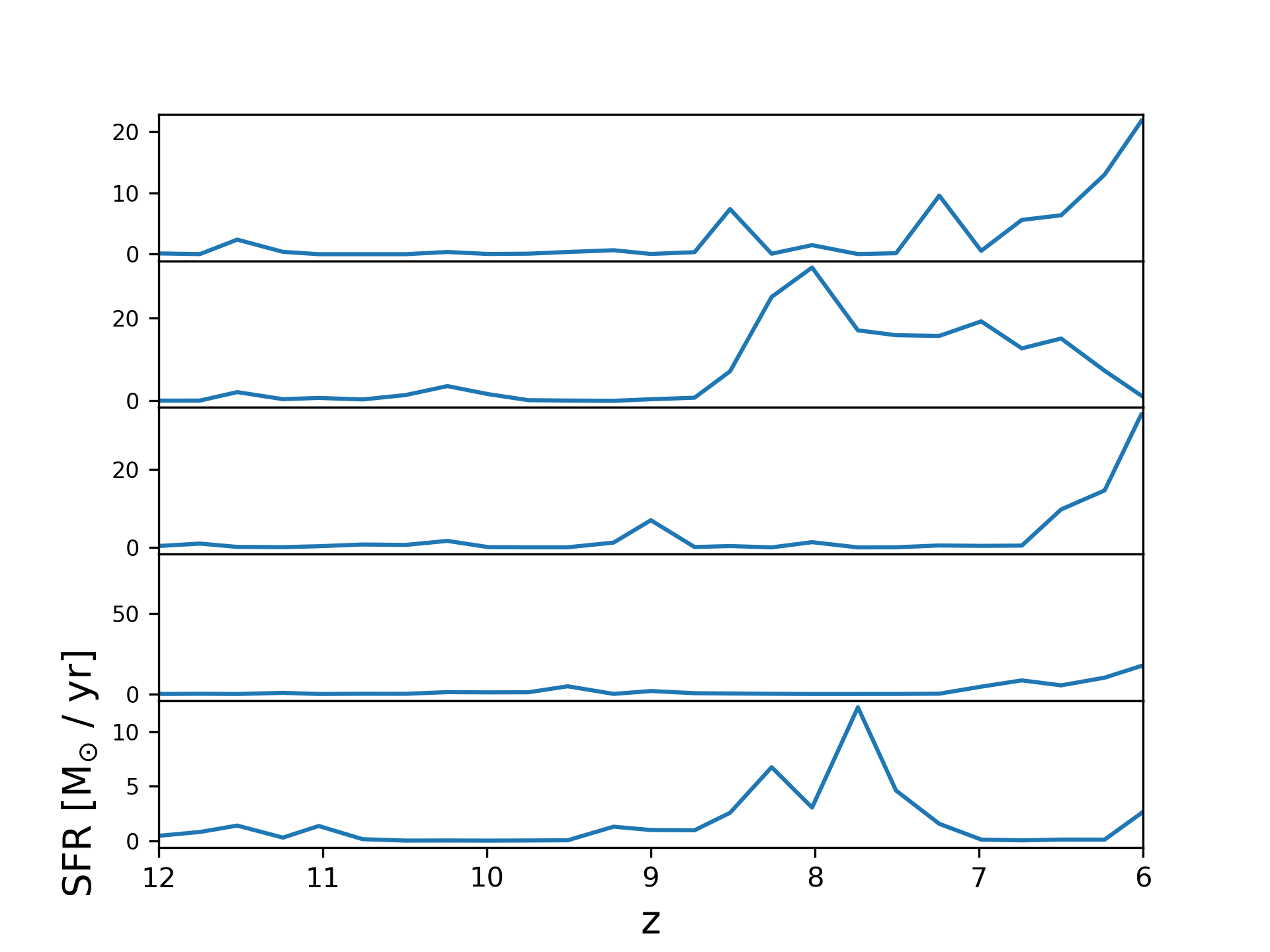}
  \caption{{\bf Model star formation histories for galaxies listed in
      Table~\ref{table:zooms}} From top to bottom, the SFHs are for
    galaxies h10,h15,h17,h20 \& h25 respectively.  \label{figure:sfh}}
\end{figure}

In this section, we aim to orient the reader around the main physical
and (mock) observable properties of our model galaxies.  We summarize
their basic physical properties in Table~\ref{table:zooms}.

We begin our tour of our model galaxies by examining their
selectability in current JWST surveys.  We present results below
$z<12$ as the dust radiative transfer is not well-resolved at higher
redshifts, owing to the relatively few simulation dust particles (which of course represent many real dust particles) constructing the
Voronoi mesh.  In the top panel Figure~\ref{figure:muv_mass_z}, we
compare the UV absolute magnitude of our model galaxies (computed at a
rest-frame wavelength of $\lambda=1500 \angstrom$) to the observed
samples of \citet{tacchella22a,austin23a,cullen24a,morales24a} and
\citet{topping24a}.\ The simulated points are color-coded by their
total dust mass.  It is clear from the top panel of
Figure~\ref{figure:muv_mass_z} that the general overlap between the
observed data and our simulated galaxy mock photometry is reasonable
at redshifts $z>10$, with typical $M_{\rm UV} \approx -18\rightarrow
-20$.  At lower redshifts ($z<10$), our galaxies are brighter than
those selected by \citet{topping24a} and \citet{morales24a}, owing to
the fact that we study individual galaxies in evolution, which get
brighter with cosmic time at these redshifts, and the relatively small volumes probed by those surveys. 

\begin{figure}
  \includegraphics[scale=0.6]{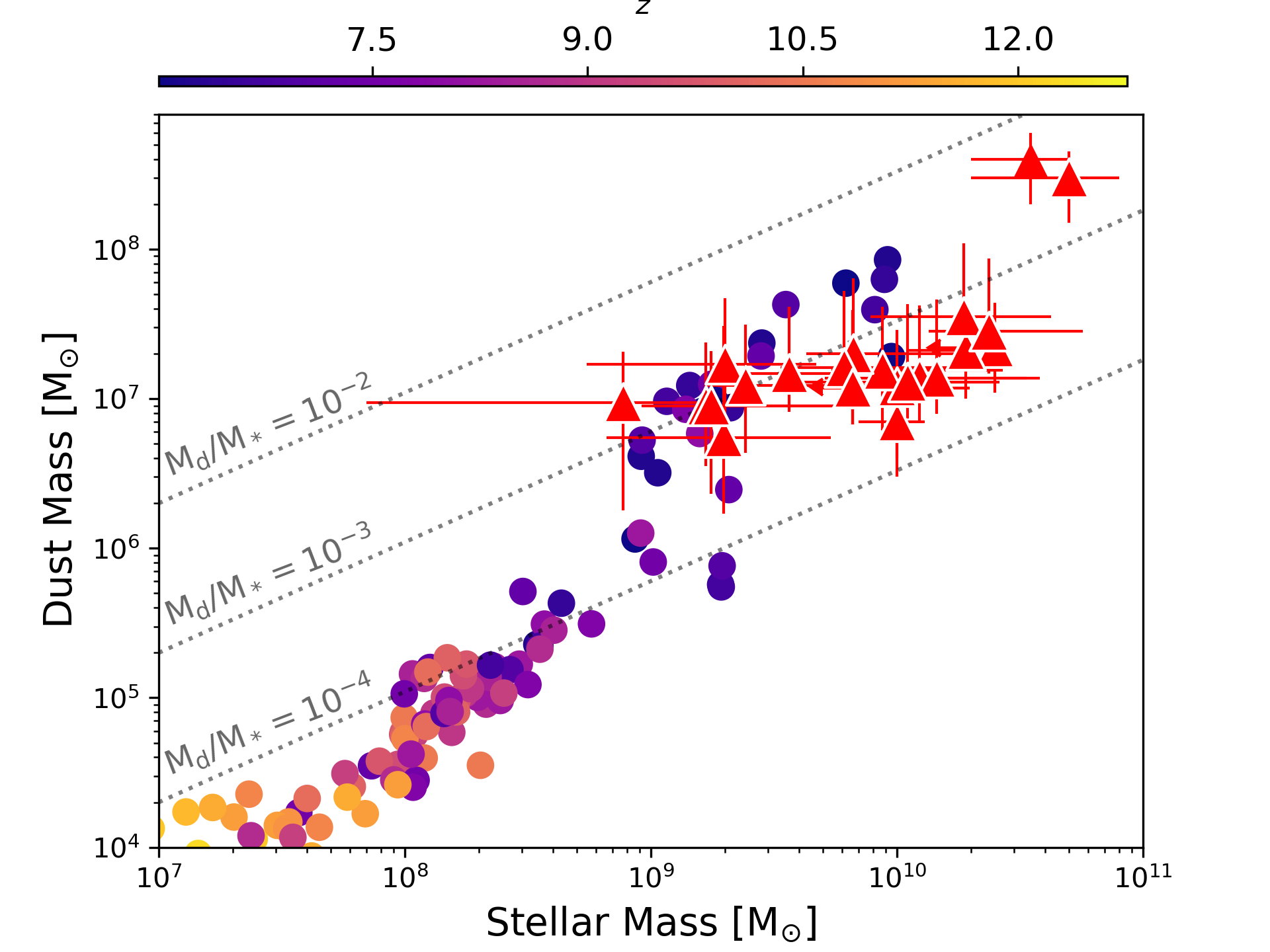}
  \caption{{\bf Comparison of $M_{\rm dust}$ vs $M_*$ for our model
      galaxies with observations demonstrates both reasonable dust properties in our model galaxies, as well as an evolutionary connection between $z>10$ UV-bright galaxies, and $z=6$ massive, infrared-luminous systems (c.f. Figure~\ref{figure:muv_mass_z}).} The circles are our simulation results at $z>6$, color-coded by their redshift, while the red triangles with errorbars are observational data taken from a range of individual detections and surveys \citep{cooray14a,watson15a,knudsen17a,laporte17a,marrone18a,hashimoto19a,bakx21a,fudamoto21a,endsley22a,dayal22a,ferrara22a,topping22a,sommovigo22a}.
\label{figure:rebels}}
\end{figure}

In the bottom panel of Figure~\ref{figure:muv_mass_z}, we show the
stellar masses of our model galaxies as a function of redshift.  In
the redshift interval of interest ($z\approx 10-12$), the stellar masses
range from $M_* \approx 10^7-10^8 M_\odot$, comparable to inferences by \citet{morales24a}.  In
Figure~\ref{figure:sfh}, we show the model star formation histories
for our $5$ model galaxies down to $z=6$.   The {\sc smuggle}
stellar feedback model drives a bursty star formation history as is
seen in other explicit feedback models \citep[e.g. {\sc
    fire;}][]{sparre17a}.

The end result for our model galaxies, which may be
detectable in the rest-frame UV at $z=10-12$ with JWST, are massive, and dust-rich systems at
$z=6$.  The individual galaxy stellar masses of our model galaxies
rise from a typical value of $M_* \approx 10^7 M_\odot$ 
 at $z=10$ to
$M_* \approx 10^{9} M_\odot$ at $z=6$.  Concomitant is the rise in
dust mass, buoyed first by production processes, followed a sharp rise
in mass owing to metal accretion in the ISM.  The rapid growth of
these galaxies renders them comparable to those detected in major ALMA surveys.  As an
example,  Figure~\ref{figure:rebels} shows the relationship
between stellar mass and total dust content of our model galaxies at
$z=6$ as compared to dust and stellar mass inferences from a range of
high-$z$ observations
\citep{cooray14a,watson15a,knudsen17a,laporte17a,marrone18a,hashimoto19a,bakx21a,fudamoto21a,endsley22a,dayal22a,ferrara22a,topping22a,sommovigo22a}.
It is evident that as galaxies in our modeled mass range approach
$z\la 8$, their dust-to-stellar mass ratios rise significantly (which is a topic we will return to in \S~\ref{section:reddened}). Figure~\ref{figure:rebels} not only highlights model
viability by demonstrating a reasonable correspondence between the
late time ($z=6$) descendants of our $z=10$ galaxy analogs, but also
demonstrates a credible evolutionary connection between UV-bright
systems detected by JWST at $z>10$, and massive, infrared-luminous systems at
$z=6$.

\section{The Role of Bursty Star Formation on Intrinsic Ultraviolet Colors}
\label{section:unreddened}

\begin{figure*}
  \centering
  \includegraphics[]{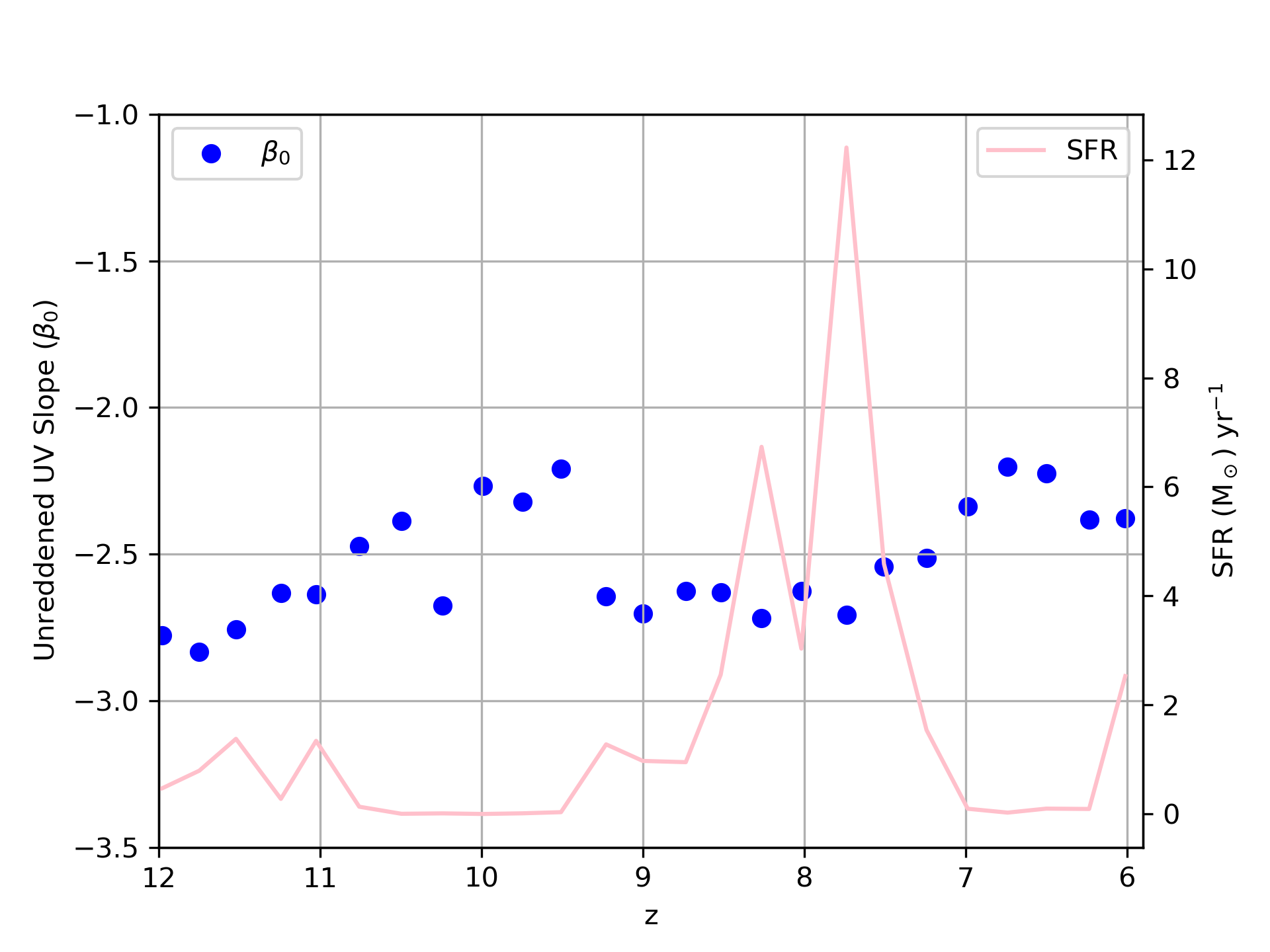}
  \caption{{\bf UV slopes from unattenuated stellar populations can show
      a diverse range of values, owing to bursty star formation
      histories in low-mass galaxies in the early Universe.}
    Unreddened $\beta_0$ (blue dots) vs $z$ for example galaxy h25.
    Aging stellar populations redden the UV slope from $z=12
    \rightarrow 9$ in this example galaxy, until bursts of star
    formation act to blue-en $\beta$.  We see a range of
    $\beta \approx [-3,-2.2]$ from intrinsic stellar populations alone (not
    including any contribution from dust or nebular continuum).  This said, as
    the stellar mass of a galaxy increases, increasingly large bursts
    of star formation are necessary to blue-en $\beta$, tamping down
    the impact of star formation below $z=7$ in this example (see
    also Figure~\ref{figure:ssfr_beta}).\label{figure:beta_z}}
\end{figure*}

\begin{figure}
  \includegraphics[scale=0.6]{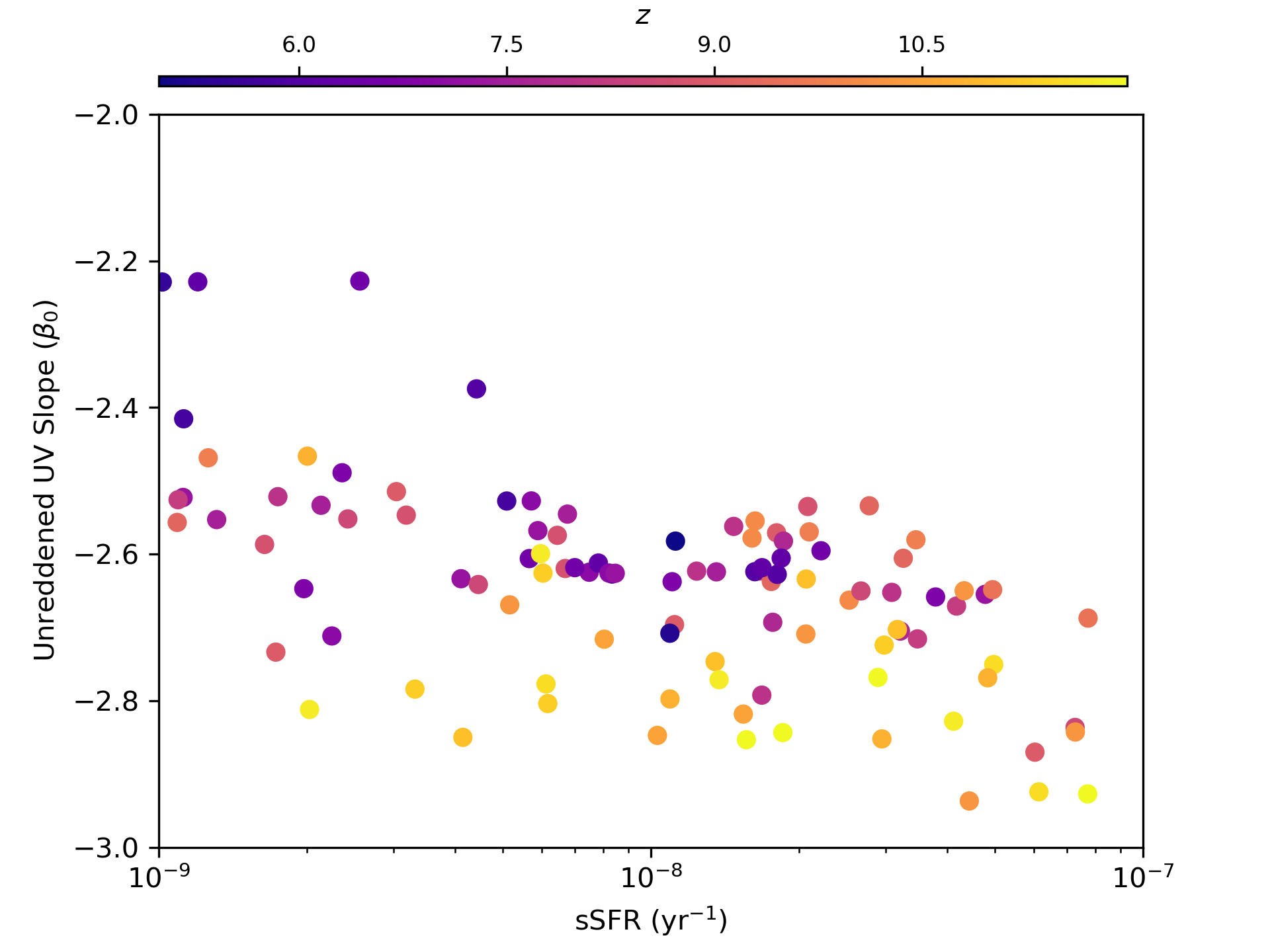}
  \caption{{\bf Relationship of galaxy  unreddened UV slope from stellar continuum alone ($\beta_0$) vs specific
      star formation rate (sSFR$\equiv$SFR/$M_*$) for all model galaxies.}  Generally, aging
    stellar populations tend to redden UV slopes, while bursts of star
    formation drive bluer UV slopes (c.f. Figure~\ref{figure:beta_z}).  This said, as the stellar mass of a galaxy
    increases, increasingly large bursts of star formation are
    necessary to blue-en the UV slope, resulting in an
    anti-correlation between $\beta_0$ and galaxy
    sSFR.\label{figure:ssfr_beta}}
\end{figure}

We now turn to our analysis of
the UV slope $\beta$ in early Universe galaxies.  In this section, our key goal is to understand the potential range of UV
slopes of unreddened galaxies, and in specific, we demonstrate
significant variability in the UV slopes of unreddened galaxies owing
entirely to bursty stellar populations.  Periods of ongoing star
formation naturally have relatively blue UV slopes, whereas periods of
star formation dormancy have redder UV slopes.

We show this point explicitly in Figure~\ref{figure:beta_z}, where we
plot the UV slope ($\beta$) for unattenuated\footnote{As a point of
methodology: all galaxy models include our full dust physics model.
To compute the unreddened UV slope in dust-free situations, we simply
neglect dust in the {\sc powderday} dust radiative transfer
calculations.} stellar populations vs redshift for an arbitrary model
galaxy in our simulation sample (h25; we choose this galaxy because
its complicated SFH provides an instructive example, though show the
equivalent plot for all model galaxies in the appendix).  We defer
discussion of the evolution of $\beta$ with redshift when dust
reddening is included to \S~\ref{section:reddened}.

Generally, the unattenuated UV slopes are at their bluest during periods
of intense star formation, and then redden during periods of low star
formation rate.  This is seen explicitly with the pink line in
Figure~\ref{figure:beta_z}, which corresponds to the right hand axis,
quantifying the galaxy star formation rate.  
As the galaxy star formation history progresses, however, the ability for a starburst to blue-en\footnote{While we know that the word ``blue-en'' does not
exist in the English language, we hereby introduce this in the same
spirit as ``redden''.} the UV slope is diminished, owing to the contribution of the underlying older stellar population to the UV colors.   This results in an inverse relationship between the UV slope and the specific star formation rate (sSFR$\equiv$ SFR/$M_*$) of a galaxy.  We quantify this in Figure~\ref{figure:ssfr_beta}, where we compare $\beta_0$
to the galaxy sSFR for all galaxy snapshots in our simulation
campaign.  Evident in Figure~\ref{figure:ssfr_beta} are both a trend with $\beta_0$ and sSFR, as well as trends at constant sSFR (i.e., vertical slices) where $\beta_0$ becomes redder at later times, reflecting the buildup of an older stellar population.


The upshot from Figures~\ref{figure:beta_z} and
\ref{figure:ssfr_beta} is that significant reddening of UV slopes in
the early Universe can occur, even prior to the onset of significant
dust obscuration.  In this particular model, we find a range from
roughly $\beta\approx -3 \rightarrow -2.2$ is attainable {\it even
  without the inclusion of dust or nebular continuum}.  Similar
effects were noted by both \citet{cullen24a} and \citet{topping24a} in
stellar population synthesis models.  
We further note that a consequence of using an explicit feedback 
model with bursty star formation histories is that the resulting galaxies are expected to experience substantial dynamic swings in their age-driven reddening.  
In contrast, galaxies modeled with smoother star formation histories -- such
as those that may be expected from cosmological simulations with an
artificial pressurization of the ISM \citep[e.g.][]{narayanan24a} -- will experience less rapid/dramatic changes in the age distribution of their stellar populations, and therefore less rapid/dynamic changes to the age-driven reddening.


\section{The Rise of Dust Obscuration at Ultra-High $z$}
\label{section:reddened}
\begin{figure*}
  \centering \includegraphics[]{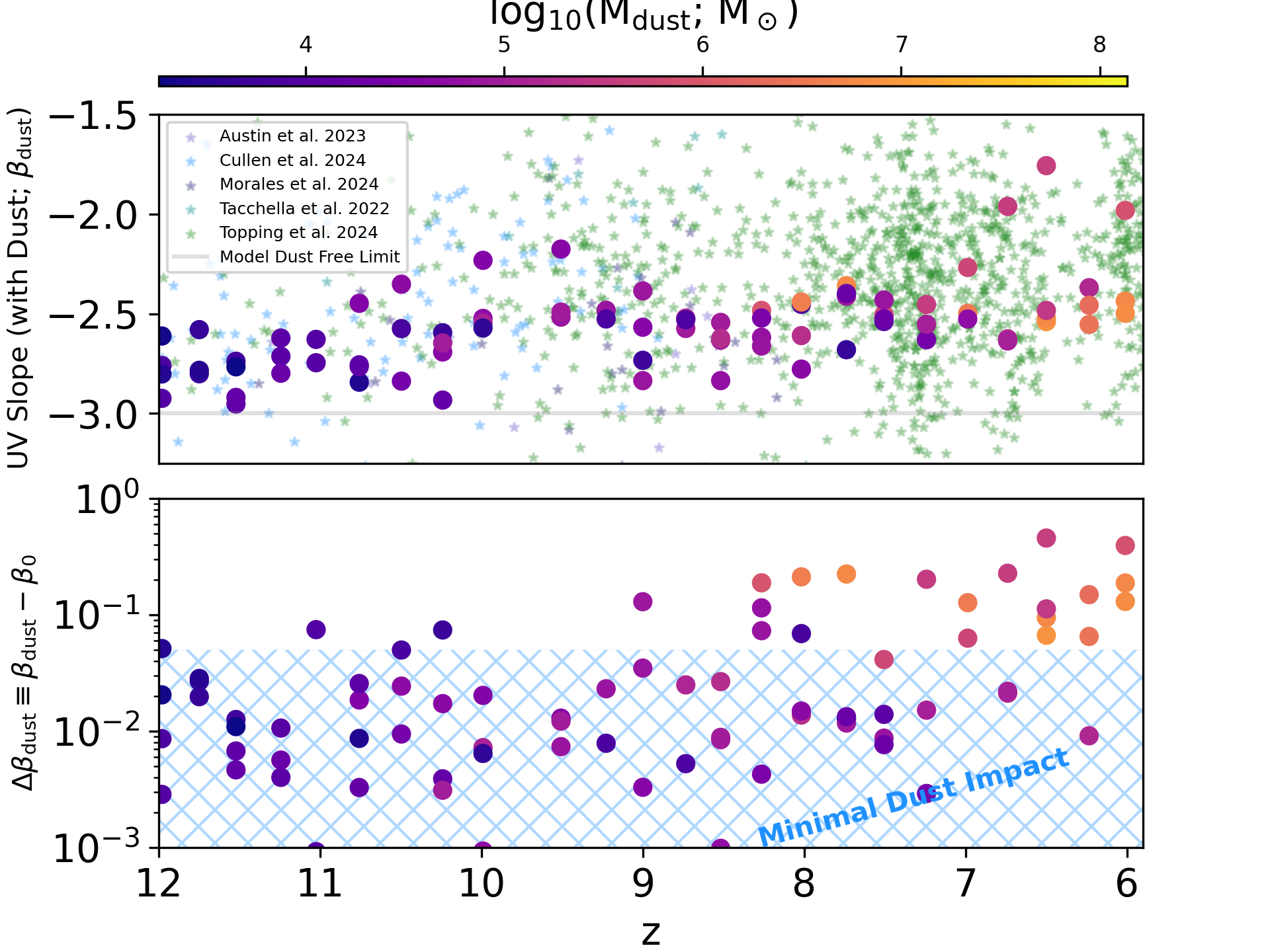}
  \caption{{\bf UV slope vs redshift for all model galaxies where we
      include the impact of dust (but not nebular continuum;
      $\beta_{\rm dust}$).}  The top panel shows this UV slope, where
    the red-purple points are our model results color-coded by the
    total dust mass, while the stars are observations from
    \citet{austin23a,cullen24a,morales24a,tacchella22a} and
    \citet{topping24a}.  The thin grey line denotes the bluest colors seen in our dust-free models. The bottom panel shows the difference in the
    intrinsic (unreddened) UV slope from the UV slope that includes
    dust ($\Delta \beta_{\rm dust}$), quantifying the impact of dust
    on the UV slope.  Generally, the increase in dust obscuration at
    lower redshifts drives redder UV $\beta$
    slopes.\label{figure:beta_delta_beta}}
\end{figure*}

Having established the typical dynamic range in $\beta_0$ owing to
bursty star formation histories (\S~\ref{section:unreddened}), we now
turn our attention to the impact of dust obscuration on the UV slope
(i.e., $\beta_{\rm dust}$).  In Figure~\ref{figure:beta_delta_beta},
we show in the top panel the UV slope $\beta_{\rm dust}$ vs $z$ for
all of our model galaxies, and in the bottom panel $\Delta \beta_{\rm
  dust} = \beta_{\rm dust}-\beta_0$. $\Delta \beta_{\rm dust}$
quantifies the direct impact of dust reddening on the intrinsic
stellar UV slope, $\beta_0$. We arbitrarily define $\Delta \beta_{\rm
  dust} < 0.05$ as the region where dust has relatively minimal
impact (this is $\sim 5\%$ of the dynamic range seen in $\beta$ from stellar populations alone), and denote this region in the bottom half of
Figure~\ref{figure:beta_delta_beta}.

\begin{figure}
  \centering \includegraphics[scale=0.6]{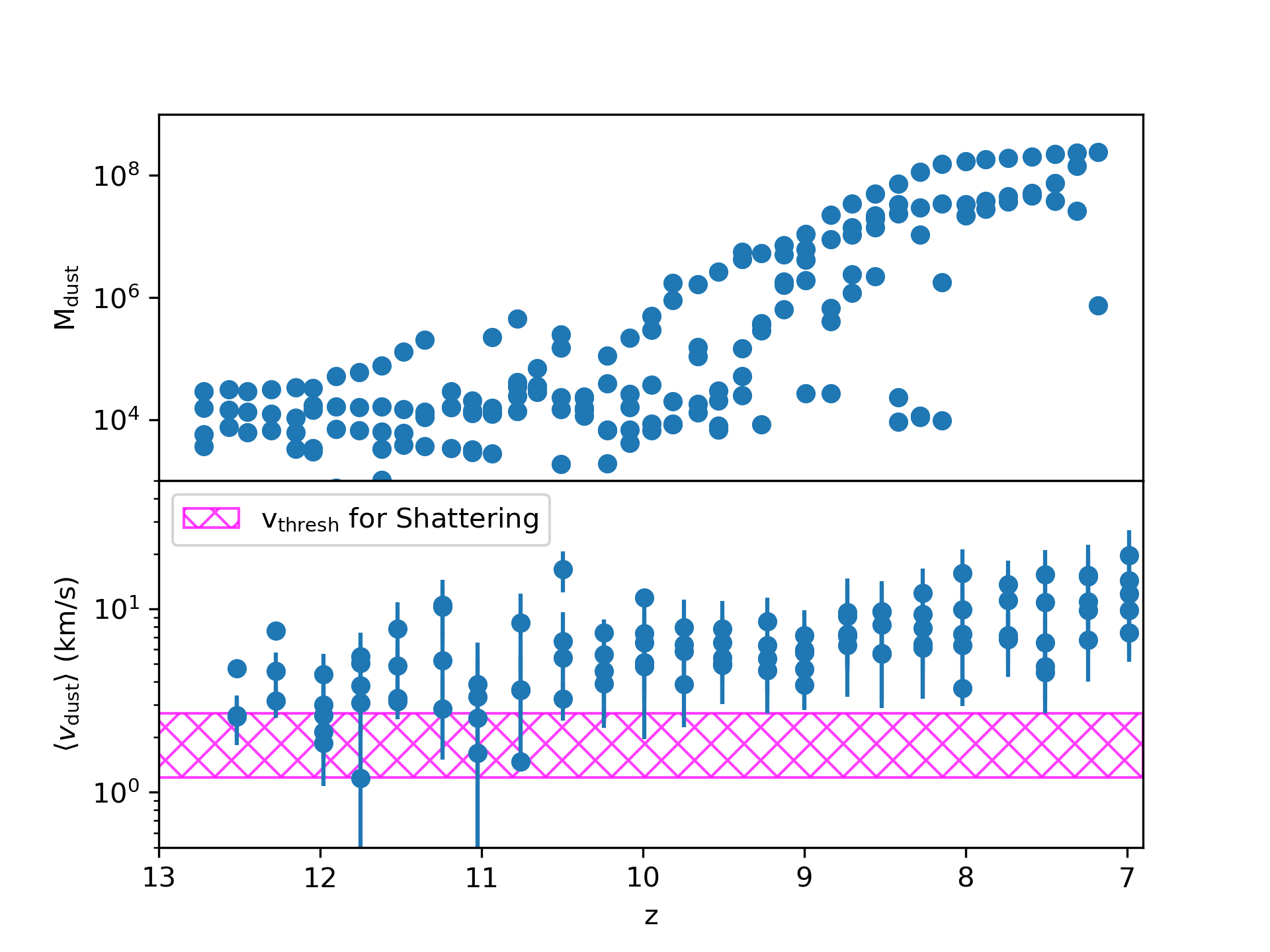}
  \caption{{\bf Evolution of dust mass and average grain velocity
      dispersions as a function of redshift.}  {\it Top:} We show the
    dust mass for all of our model galaxies as a function of redshift.
    {\it Bottom:} This rapid dust growth at $z\approx10$ is due to the
    production of small dust grains via grain-grain shattering events.
    These events increase with the velocity of grain collisions.
    Here, we plot the median relative velocity between two dust grains for each
    model galaxy in our simulation sample.  The error bars show the
    $1\sigma$ dispersion in these velocity dispersions, and the pink
    shaded region shows the range of velocity (species-dependent)
    thresholds employed for shattering to occur in grain-grain
    collisions. \label{figure:mdust_z}}
\end{figure}

\begin{figure}
  \centering\includegraphics[scale=0.6]{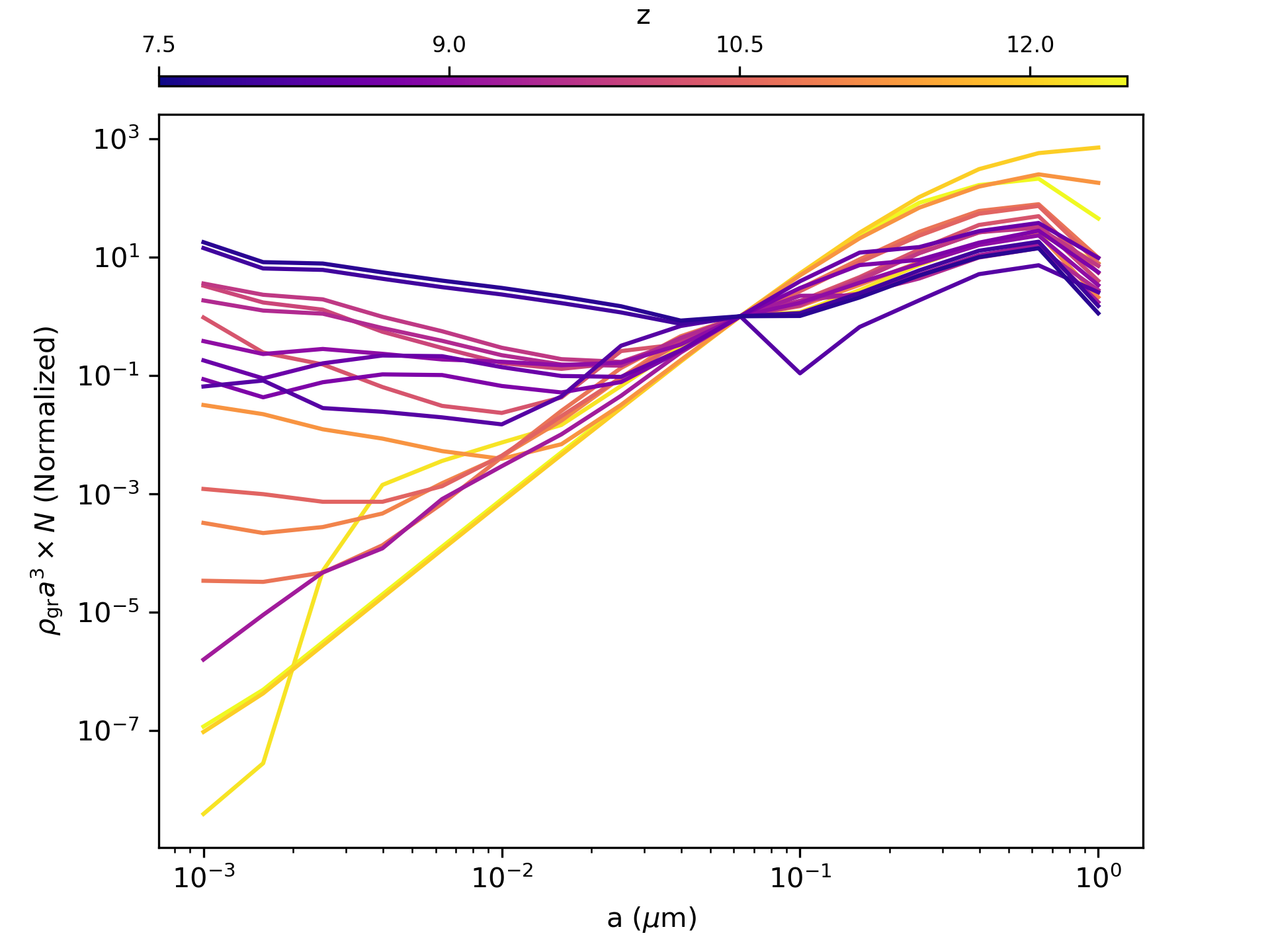}
  \caption{{\bf The relative fraction of small dust grains increases
      with decreasing redshift, resulting in rapid dust growth
      (c.f. Figure~\ref{figure:mdust_z}).} Shown is the normalized
    grain size distribution for model galaxy h25 (as an example),
    color-coded by redshift.  The size distributions are normalized at
    $a\approx 0.06 \, \mu$m, the canonical dividing line between 'small'
    and 'large' grains in the literature.  Shattering events drive
    increased fractions of small grains, which enhances accretion of metals due to the large surface area-to-mass ratio of small dust grains. \label{figure:gsd}}
\end{figure}

Generally, at $z \approx 8-9$, the galaxy population that we model
transitions from having relatively minimal impact of dust on the UV
slope to more of an impact, with approximately half of the galaxies
transitioning above $\Delta \beta_{\rm dust} > 0.05$ by $z=8$.  By $z=6-8$, dust begins to impact a substantial fraction of galaxy UV slopes.
This corresponds to an overall trend in $\beta_{\rm dust}$ to rise
with decreasing redshift, with comparable UV slopes as the
observational comparison sets of \citet{cullen24a} and
\citet{topping24a}. In general, between $z=12\rightarrow6$, we see a
typical rise from our dust-free solution of $\beta_0 \approx -3$ to as red as 
$\beta_{\rm dust} \approx -2$. This overall rise
in $\beta$ with $z$ is not necessarily monotonic, however, owing to
bursts of star formation that can somewhat blue-en $\beta_{\rm dust}$
(c.f. \S~\ref{section:unreddened}). We note that the impact of
  these bursts on blue-ening the UV SED may be mitigated if nebular
  continuum emission plays a significant role in the UV SEDs, and we
  discuss this further in \S~\ref{section:nebular}.

The increase in dust reddening at $z \approx 8-9$ owes to a sudden
increase in the dust masses at this time.  In the top panel of
Figure~\ref{figure:mdust_z}, we plot the dust mass evolution for all of our
model galaxies (this information was otherwise available in
Figure~\ref{figure:rebels} as well, and is just distilled slightly
differently here).  At $z \approx 10$, we see rapid growth in galaxy
dust masses for the mass range considered in our simulation sample,
with an increase by a factor $100-1000$ by $z \approx 6$.  This
allows the relatively blue galaxies at high-$z$ to transition into
dusty star-forming galaxies with dust masses comparable to
ALMA-detections at $z \approx 6$ (Figures~\ref{figure:muv_mass_z} and
~\ref{figure:rebels}).  This transition to an epoch of dust mass
growth owes to the production of small grains in grain-grain
shattering events enabled by a turbulent ISM associated with star
formation.

In the bottom panel of Figure~\ref{figure:mdust_z}, we show the
average velocity between any two simulation dust particles for every galaxy modeled
in our sample, with error bars denoting the $1\sigma$
dispersion in each galaxy snapshot.  The pink shaded region shows the
imposed velocity thresholds (that are species-dependent) for grain
collisions to result in a shattering event: collisions above this
threshold velocity result in a grain-grain shattering, and the
increased production of small grains.  As the galaxy star formation
rates rise with time, the average interstellar velocity dispersion
increases, and subsequently so does the grain-grain shattering rate.
The surface areas per mass are larger for small dust grains than large
ones, and the growth rates via metal
accretion are faster\footnote{This is why when you want to cool down the drink that
you've prepared to enjoy while reading this paper, crushed ice will do the job faster than the fancy large cubes, given its larger surface area per unit mass, though at the peril of totally watering the drink down.}  (c.f. Equation~\ref{equation:growth}).  This is
quantified even further in Figure~\ref{figure:gsd}, where we show the
normalized dust grain size distribution evolution with redshift for
example galaxy h25; the size distributions are normalized at
$a=0.06\,\mu$m, the canonical dividing line between small and large
grains \citep[though this choice is arbitrary;][]{hirashita15a}.  It is clear that the
small-to-large ratio increases dramatically over the redshift range
considered. As a result, the epoch of significant dust growth is
associated with the epoch of significant rises in galaxy star
formation histories for a given galaxy.  These growth rates eventually plateau as the dust spatial distribution becomes more diffuse as the galaxy grows, and the metal densities decrease.  For the range of masses
modeled here, the epoch of peak dust growth during the redshift range considered is approximately $z\approx8-10$.

\section{On The Role of Nebular Continuum on the UV Slopes of Early Galaxies}
\label{section:nebular}
HII regions surrounding massive stars can redden the UV continuum
owing to free-free and free-bound processes \citep{byler17a,katz24a}.  Indeed,
\citet{stanway16a,topping22a,topping24a} and \citet{cullen24a} all
find that intrinsic stellar populations with nebular continuum (but no
dust) can achieve minimum colors of $\beta \approx -2.6$.  In
this section, we investigate the impact of nebular continuum on our
model galaxies.  

\begin{figure}
  \centering \includegraphics[scale=0.6]{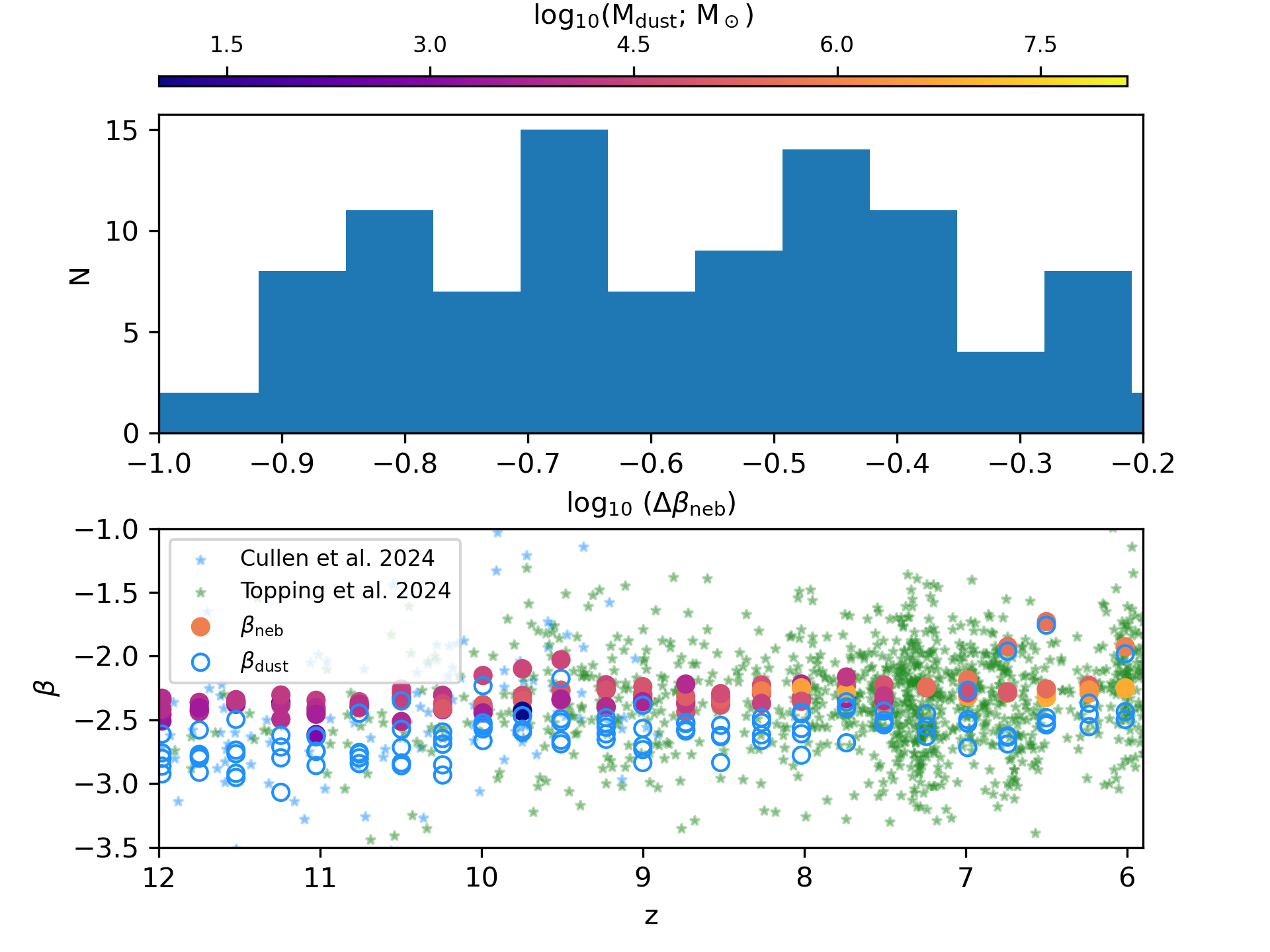}
  \caption{{\bf Nebular Continuum drives a Reddening of the UV slope
      of $\beta \approx 0.2-0.4$, but results in galaxies that are too red
      at very high redshifts ($z>10$). }  {\it Top:}  Histogram of
    $\Delta \beta_{\rm neb}$ (i.e., the difference in UV slope
    including dust and nebular continuum with that of just including
    dust) for all of our model galaxies.  {\it Bottom:} Evolution of
    $\beta_{\rm neb}$ (i.e., including dust and nebular continuum)
    with redshift (colored points), compared to the UV slope just
    including dust ($\beta_{\rm dust}$: empty blue circles).  When including nebular
    continuum, the colors are too red to match the highest-$z$
    observed data ($z>10$). We expand on possible physical
    explanations for this discrepancy in
    \S~\ref{section:discussion}. \label{figure:beta_z_neb}}
\end{figure}

In the top panel of Figure~\ref{figure:beta_z_neb}, we show a
histogram of $\Delta \beta_{\rm neb}$ (i.e., isolating the impact of
nebular reddening over that of dust) for all of our model galaxies,
across redshift.  The distribution is broad, but peaks in a region $\Delta
\beta_{\rm neb} \approx 0.2-0.4$, quantifying the typical reddening
we expect in our models from nebular continuum emission.

This said, we argue that the contribution of nebular continuum to the
UV continuum slope cannot be uniform across redshift.  In the bottom
panel of Figure~\ref{figure:beta_z_neb}, we show the evolution of
$\beta_{\rm neb}$ (i.e. $\beta$ including dust and nebular continuum)
with redshift, and as a reference, in the light colored blue circles,
show $\beta_{\rm dust}$ (i.e. the UV slope suffering from dust
reddening alone).  At the earliest times ($z>10$), the model with
nebular continuum cannot produce UV slopes sufficiently blue to match
the extremely blue ($\beta < -2.7$) observed slopes.  At the same
time, the model with only dust obscuration {\it can} result in
sufficiently blue colors at $z>11$ to compare well against
observational constraints, owing to the relative lack of dust and
the dominance of intrinsic stellar populations in the UV SED.  One
possible inference from this result is an increasing escape fraction
of ionizing photons with increasing redshift.  We expand on this
possibility further in \S~\ref{section:discussion}.

\section{Discussion}
\label{section:discussion}


\subsection{Comparison with Observational Studies}
JWST has allowed for a significant number of detections of galaxies at ever-higher
redshifts, allowing for a direct comparison between our model results
and observations at $z>10$.  While individual studies do
not (yet) converge on the magnitude of the trend of $\beta_{\rm dust}$
with redshift, in aggregate there are clear observational trends
demonstrating the reddening of the UV slopes of galaxies between
$z=12\rightarrow6$.  Of course any such trend would be expected to be
mass dependent: more massive halos collapse earlier, begin star
formation earlier, and begin to form dust earlier.  This said, because
our model sample of galaxies overlaps with the rough UV luminosity
range of those observed (Figure~\ref{figure:muv_mass_z}), as well as
at least  some stellar mass constraints \citep{morales24a}, our modeled trend of
$\beta_{\rm dust}$ vs $z$ appears to be in reasonable correspondence with that
observed 
\citep[Figure~\ref{figure:beta_z}; also][]{tacchella22a,austin23a,cullen24a,morales24a,topping24a}.  As
observations push to deeper limits (or, for example, probe lower
intrinsic masses via the aid of gravitational lensing), one may expect
bluer UV slopes to persist even below $z<10$.

What is less clear, however, is whether the correspondence between our
model $\beta_{\rm dust}$ and $z$ and observed galaxies owes in reality
to the buildup of galactic dust, as we have suggested.  We have
painted a circumstantial picture that links the buildup of dust in the
current $M_{\rm UV} \approx -18-20$ JWST detections at $z>10$ to ALMA
detectable galaxies at $z \approx 6$, though in reality there is little direct
observational evidence for the emergence of cosmic dust in these
galaxy populations over this redshift range.  The stellar
age-reddening degeneracy prevents this from being fully deducible from
UV continuum observations alone (c.f. \S~\ref{section:unreddened}). Put directly, in our models, we would attribute at least some of the reddening observed in $\beta$ through $z=9$ by \citet{cullen24a} and \citet{topping24a} as being due to stellar aging.  In order to distinguish between this scenario vs early dust buildup, we advocate for two
potential observational routes to directly test our bespoke
assertions.  The first is the pursuit of thermal infrared emission in
$z>10$ galaxies, which may require the assistance of gravitational
lensing given current instrument sensitivities.  The second is via the
inference of cosmic dust via the $2175 \AA$ absorption feature
\citep[e.g.][]{witstok23a,markov24a}.  This UV ``bump''
 potentially owes to ultrasmall carbonaceous dust
grains, and may therefore betray the emergence of cosmic dust in
galaxies.  If this feature is related to PAHs, and if PAHs form in a
top-down shattering formation scenario as is suggested by at least
some models \citep[e.g.][]{narayanan23a}, then the inference of the
  $2175 \AA$ bump may additionally provide evidence for a rapid growth
  scenario that owes to grain-grain shattering, and the associated
  increased growth rates on small dust grains in the turbulent ISM of
  high-$z$ galaxies (e.g. Figure~\ref{figure:mdust_z}).

  We now turn our discussion to the implications of very blue UV SEDs
  ($\beta_{\rm dust} \approx -3$), in light of the potential reddening
  effects of nebular continuum from {\sc HII} regions around massive
  stars.  As we demonstrated in Figure~\ref{figure:beta_z_neb}, a
  reasonable model for the nebular continuum contribution to the
  reddening of UV slopes at high-$z$ results in the complete lack of
  ability to produce the very blue galaxies observed by JWST at $z>10$
  \citep[e.g.][]{cullen24a,topping24a}.  As a reminder, in our model,
  we assume $f_{\rm esc} = 0$ for all {\sc HII} regions.  Therefore,
  one possible implication of Figure~\ref{figure:beta_z_neb} is an
  evolution of $f_{\rm esc}$ with redshift, such that the emission
  star-forming regions in galaxies at increasingly high redshift are
  more easily able to escape their {\sc HII} regions.  Indeed, there
  may be some observational evidence for such a heuristic model.
  \citet{topping24a} performed {\sc cloudy} photoionization modeling
  of {\sc HII} regions, and demonstrated that large escape fractions
  may be necessary to reproduce UV colors as blue as has been observed
  by both their group, as well as \citet{cullen24a}.  This is
  consistent with the weak emission line spectra observed by
  \citet{topping24a} in their sample.  Beyond this, \citet{chisholm22a}
  demonstrated a strong correlation between $\beta$ and $f_{\rm esc}$
  for low-redshift ($z\approx0.3$) galaxies.  By applying the
  \citet{chisholm22a} $\beta-f_{\rm esc}$ relation to their observed
  sample, \citet{cullen24a} infer a potential rise of up to a factor
  $\sim 3$ in $f_{\rm esc}$ for galaxies binned below and above $z =
  10.5$ \citep[though see ][which suggests lower $f_{\rm esc}$ values may be necessary to match ionizing photon count constraints]{munoz24a}.   This said, we note that it is possible that observed galaxies with very blue UV slopes may be undergoing small bursts of star formation that drive very blue colors.  As demonstrated via population synthesis modeling by \citet{topping24a}, $5$ Myr after a burst, the nebular continuum emission is weakened sufficiently that for a short period the UV colors can be extremely blue ($\beta \approx -2.8$) even for $f_{\rm esc} = 0$ models.   This effect may not be captured in our models, given the cadence of our snapshot output.



  \subsection{Relationship with Existing Theoretical Models}
  The observation of very blue UV-bright galaxies discovered by JWST
  has generated a flurry of theoretical activity in recent years.
  Much of the theoretical effort has framed the problem in terms of
  the relative non-evolution of the UV luminosity function above $z>7$
  \citep[e.g.][]{naidu22a,bouwens23a,donnan23a,finkelstein23b,donnan23b,harikane23a,perezgonzalez23a}.
  Here, one natural solution is the reduction of dust attenuation
  at $z>7$, to offset the natural decline in massive halos with
  increasing redshift.  \citet{ferrara23a} developed an analytic model
  for the observed galaxy abundances and UV luminosity functions at
  high-redshift.  An important aspect to
  this model, as well as others that aim to model dust at very high
  redshift \citep[e.g.][]{zhao24a} is that they are constrained by the
  boundary conditions of relatively little dust attenuation at $z>10$,
  as well as the inferred presence of significant dust reservoirs from
  ALMA detections of massive galaxies at $z\sim6$.  \citet{ferrara23a}
  demonstrate that a decrease in dust attenuation at $z>11$ can
  satisfy observational constraints on the UV luminosity function,
  which is a coupled concept with producing a population of extremely
  blue galaxies.  These authors are agnostic as to the origin of the
  lack of dust attenuation, positing possibilities such as dust
  removal due to early winds, offset star-dust geometries
  \citep{ziparo23a}, as well as low dust to stellar mass ratios.  Our
  interpretation, based on the cosmological hydrodynamic zoom-in
  simulations performed here, supports the latter scenario.

  The scenarios proposed above are fundamentally different: in two of
  the physical models (early dust ejection, as well as complex
  star-dust geometries), the premise is that the dust is
  already in place at early times, but is simply low opacity.
  This is in contrast to a model in which galaxies -- comparable to
  the mass currently being detected by JWST at $z>10$ -- grow
  significant dust reservoirs during this transition epoch.  Our model
  clearly advocates for the latter scenario.  By and large,
  simulations that aim to explicitly model the evolution of dust
  masses during the Epoch of Reionization agree that significant dust
  growth occurs in galaxies of order the mass range modeled here
  between $z>10$ and $z \approx 6$
  \citep{graziani20a,dayal22a,lower23a,esmerian23a,lewis23a,choban24b}.  This
  said, models differ in detail regarding the origin of this dust
  (i.e., growth-dominated vs production dominated), the magnitude of dust growth, and when
  the transition from production dominated to growth dominated occurs.
  Sorting out the origin of these differences is complicated as there
  is a degeneracy between differences in the underlying algorithms for
  dust evolution and the ISM physical conditions in varying
  hydrodynamic models, and would benefit from targeted numerical
  experiments and code comparison studies comparable to the {\sc
    agora} simulation suite \citep{kim14a}.

A handful of studies aimed at probing $\beta$ in the context of galaxy simulations and dust find similar levels of reddening in the UV SED owing to the contribution of dust. \citet{esmerian23a} and \citet{smith23a} implemented a single grain-size dust model into {\sc ART} and {\sc RAMSES}, respectively, and found a similar range of $\beta$ values reddening owing to dust at $z<9$.  \citet{katz23a} find even redder colors when considering $\beta_{\rm neb}$, though implement a more simplified dust model with an SMC-type extinction curve.

Finally, on the front of the contribution of nebular continuum to
$\beta$, \citet{katz24a} presented a recent series of numerical
experiments, studying the impact of nebular continuum on the UV
spectra of early Universe galaxies.  \citealt{katz24a} demonstrate a
reddening of $\Delta \beta_{\rm neb} \approx 0.6$ (going from $\beta_0
\approx -3.1 \rightarrow -2.5$ in the most extreme case), similar in magnitude to
our more simplified models for nebular emission.    Even when considering the impact of IMF variations and the assumed density of nebular regions, \citet{katz24a} finds a minimum slope of $\beta_{\rm neb} \approx -2.7$, just slightly bluer than our study.

\section{Summary}
\label{section:summary}
In this paper, we have combined a suite of cosmological zoom-in
simulations designed to represent massive galaxies in the early
Universe with dust radiative transfer modeling to study the impact of
bursty star formation histories, the production and growth of dust,
and emission from nebular regions on the rest-frame UV continuum slope
($\beta$) in galaxies.  We designed our study as a series of numerical
experiments, by adding more physics in the radiative transfer modeling
successively, and studying the results.  In particular, we simulated
the UV slopes of (i) unreddened populations within simulated
high-redshift galaxies, (ii) reddened populations by adding the impact
of dust (in radiative transfer), and (iii) reddened populations with
the contribution of nebular continuum.

Our main results follow.

\begin{enumerate}
\item We simulate galaxies with $z=0$ parent halo masses $\sim 2-6
  \times 10^{13} M_\odot$, which results in galaxies with $z=6$
  stellar masses $M_* \approx 0.3-2 \times 10^{10} M_\odot$. Galaxies
  of this mass have UV luminosities similar to those being detected in
  JWST surveys at $z>10$ (Figure~\ref{figure:muv_mass_z}).  We
  therefore use these galaxies to study the impact of bursty star
  formation histories, dust, and nebular continuum on $\beta$:

  \item {\bf Unreddened stellar populations} exhibit a diverse range of
    intrinsic UV slopes, with values ranging from a dust-free $\beta_0
    \approx -3$ to values as red as $\beta_0 \approx -2.2$
    (Figure~\ref{figure:beta_z}).  The somewhat red values of $\beta_0$,
    even without the reddening effects of dust or nebular continuum, owe to long delays
    between bursts in bursty star formation histories
    (Figure~\ref{figure:sfh}).  Bursts of star formation counter-act
    this reddening, though are less effective as the underlying older
    stellar population grows. As a result, there is  an
    inverse correlation between the intrinsic UV slope and sSFR for
    early galaxies (Figure~\ref{figure:ssfr_beta}).

    \item {\bf When considering the effect of dust}, the UV colors
      galaxies become successively redder between $z=12 \rightarrow 6$
      (Figure~\ref{figure:beta_delta_beta}), due to the rapid growth
      of dust mass during this period (Figure~\ref{figure:mdust_z}).
      This growth in dust is due to grain-grain shattering in a
      turbulent ISM, and enhanced dust growth rates on small dust
      grains due to their larger surface areas
      (Figures~\ref{figure:mdust_z} and ~\ref{figure:gsd}).  $z
      \approx 8-9$ is roughly where galaxies transition from being
      unaffected by dust to affected by dust in the UV (Figure~\ref{figure:beta_delta_beta}), though this is true only
      for galaxies within the mass range that we simulate.  This suggests that the reddening of UV colors seen in recent JWST observations during this epoch owes to the buildup of the first major dust reservoirs in these galaxies.

\item {\bf The inclusion of nebular continuum} reddens the UV slope by
  a median factor $\Delta \beta_{\rm neb} \approx 0.2-0.4$.  This
  said, when including nebular continuum, our highest redshift
  galaxies ($z\approx12$) are insufficiently blue compared to
  observations, and we cannot achieve the dust free-like blue UV
  colors of $\beta \approx -3$ as is observed
  \citep{cullen24a,topping24a}.  This may imply an evolving escape
  fraction from HII regions with redshift, such that the highest
  redshift sources have significant escape fractions.

\end{enumerate}

\section{Acknowledgements}
Desika, Dan and Steve would all like to express appreciation to
Adriano Fontana, Paola Santini, Jim Dunlop, Davide Elbaz, Alice
Shapley, and Rachel Somerville for having organized ``The Growth of
Galaxies in the Early Universe - IV'' at the Sexten Center for
Astrophysics in 2024.  This project idea came from talks at this
conference, followed by conversations had during a night out in Venice
on the way home.  Desika would like to additionally thank Sidney Lower
for having left behind enough legacy code from her PhD thesis at the
University of Florida in an easily digestible enough format that the
analysis for this paper was fairly painless.  Desika is grateful to
the Institute for Astronomy at the University of Edinburgh, and the Cosmic Dawn Centre at the Niels Bohr Institute for
graciously hosting him while he worked on significant parts of this
paper, and thanks Gabe Brammer, Romeel Dav\'e and Harley Katz for helpful
conversations while preparing this manuscript.  DN and PT were funded by NASA ATP grant 80NSSC22K0716, and
are grateful to NASA for trusting us enough to fund what probably seemed like a fairly
risky science idea of modeling active dust in galaxies.  DN is additionally grateful to the Aspen Center for
Physics, which is supported by NSF grant PHY-1607611, which is where
the original framework for {\sc powderday} was envisioned. F. Cullen acknowledges support from a UKRI Frontier Research Guarantee Grant (PI Cullen; grant reference EP/X021025/1). FM acknowledges funding by the
European Union - NextGenerationEU, in the framework of the
HPC project – “National Centre for HPC, Big Data and Quan-
tum Computing” (PNRR - M4C2 - I1.4 - CN00000013 – CUP
J33C22001170001).


\bibliographystyle{mnras}
\bibliography{./full_refs}

\appendix Here we show the relationship between $\beta_0$, redshift, and
SFR for all galaxies in our model sample (aside from h25, which is
already shown in Figure~\ref{figure:beta_z}).

\begin{figure*}
  \centering
  \includegraphics{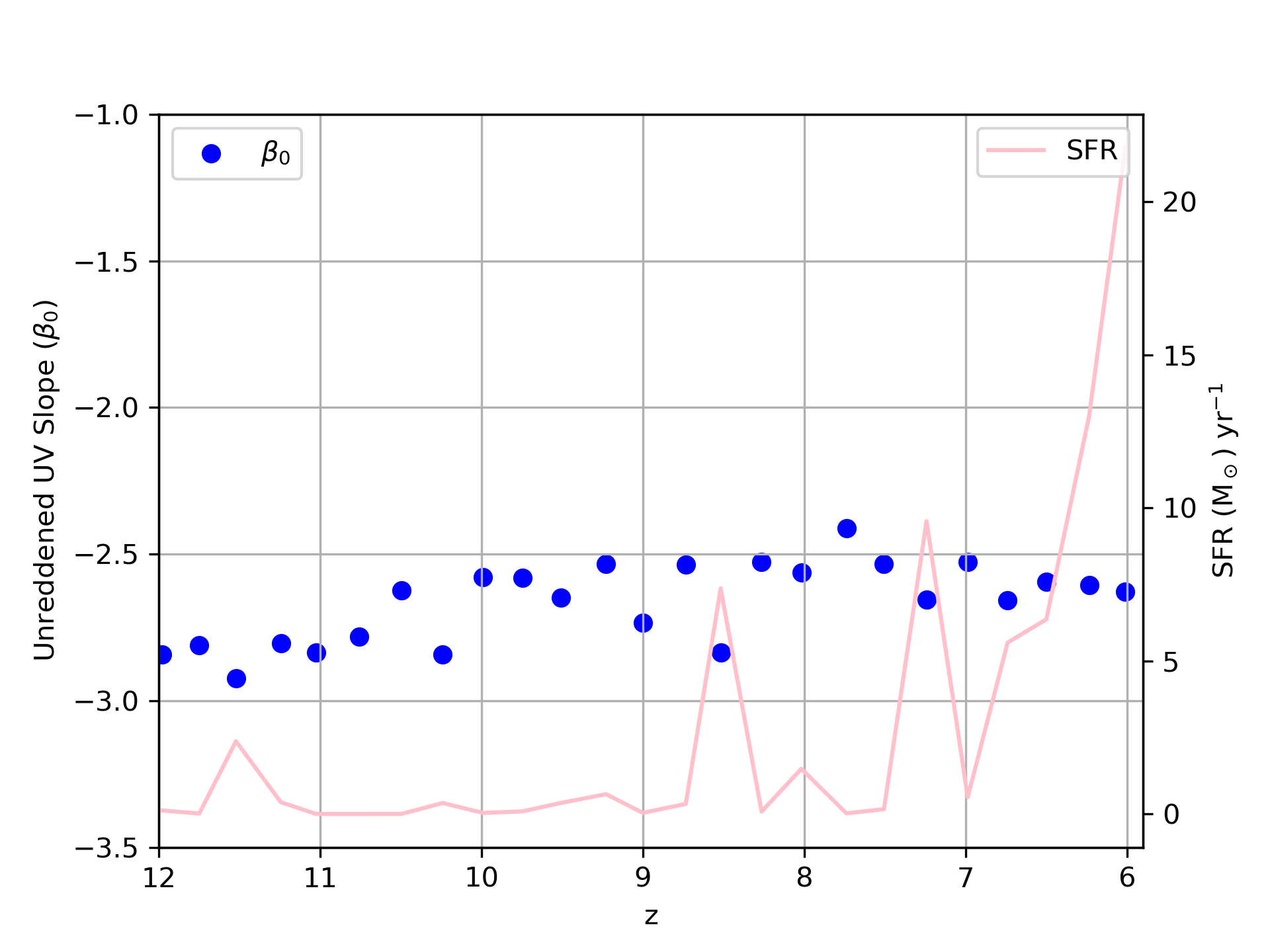}
  \caption{Unreddened UV slope ($\beta_0$) vs $z$ for galaxy h10.  See Figure~\ref{figure:beta_z}, and the associated text, for details.}
\end{figure*}

\begin{figure*}
  \centering
  \includegraphics{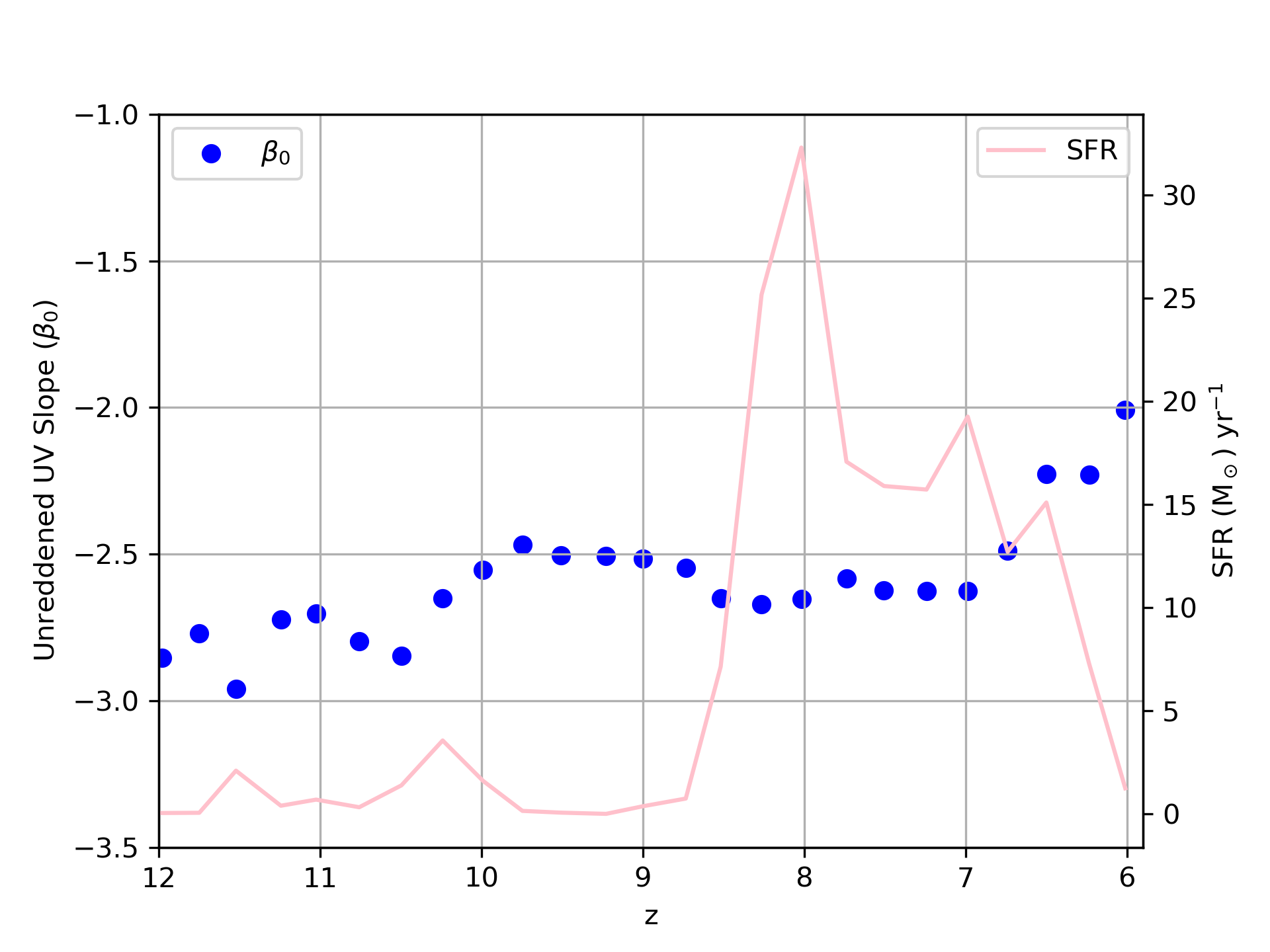}
  \caption{Unreddened UV slope ($\beta_0$) vs $z$ for galaxy h15.  See Figure~\ref{figure:beta_z}, and the associated text, for details.}
\end{figure*}

\begin{figure*}
  \centering
  \includegraphics{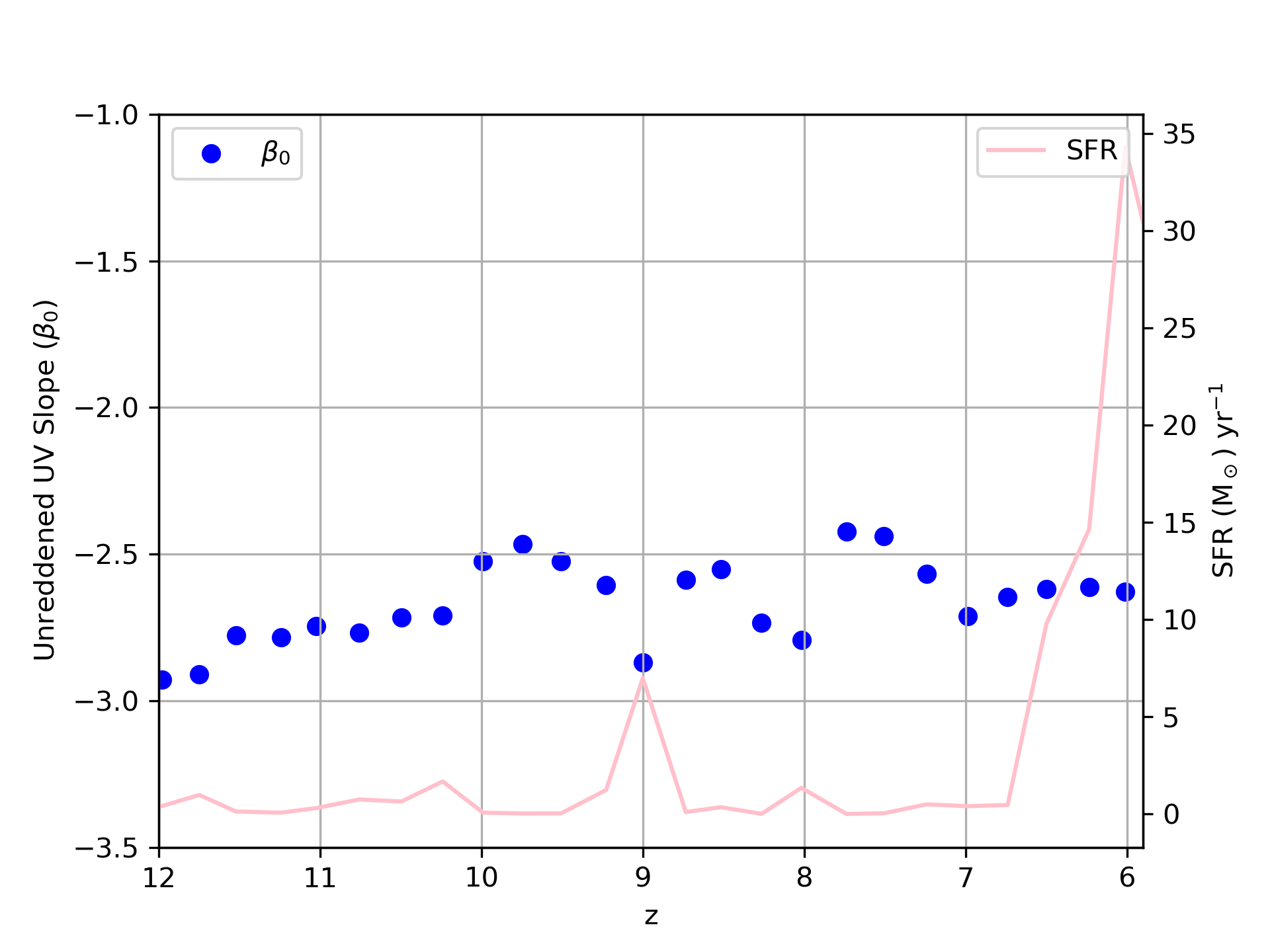}
  \caption{Unreddened UV slope ($\beta_0$) vs $z$ for galaxy h17.  See Figure~\ref{figure:beta_z}, and the associated text, for details.}
\end{figure*}

\begin{figure*}
  \centering
  \includegraphics{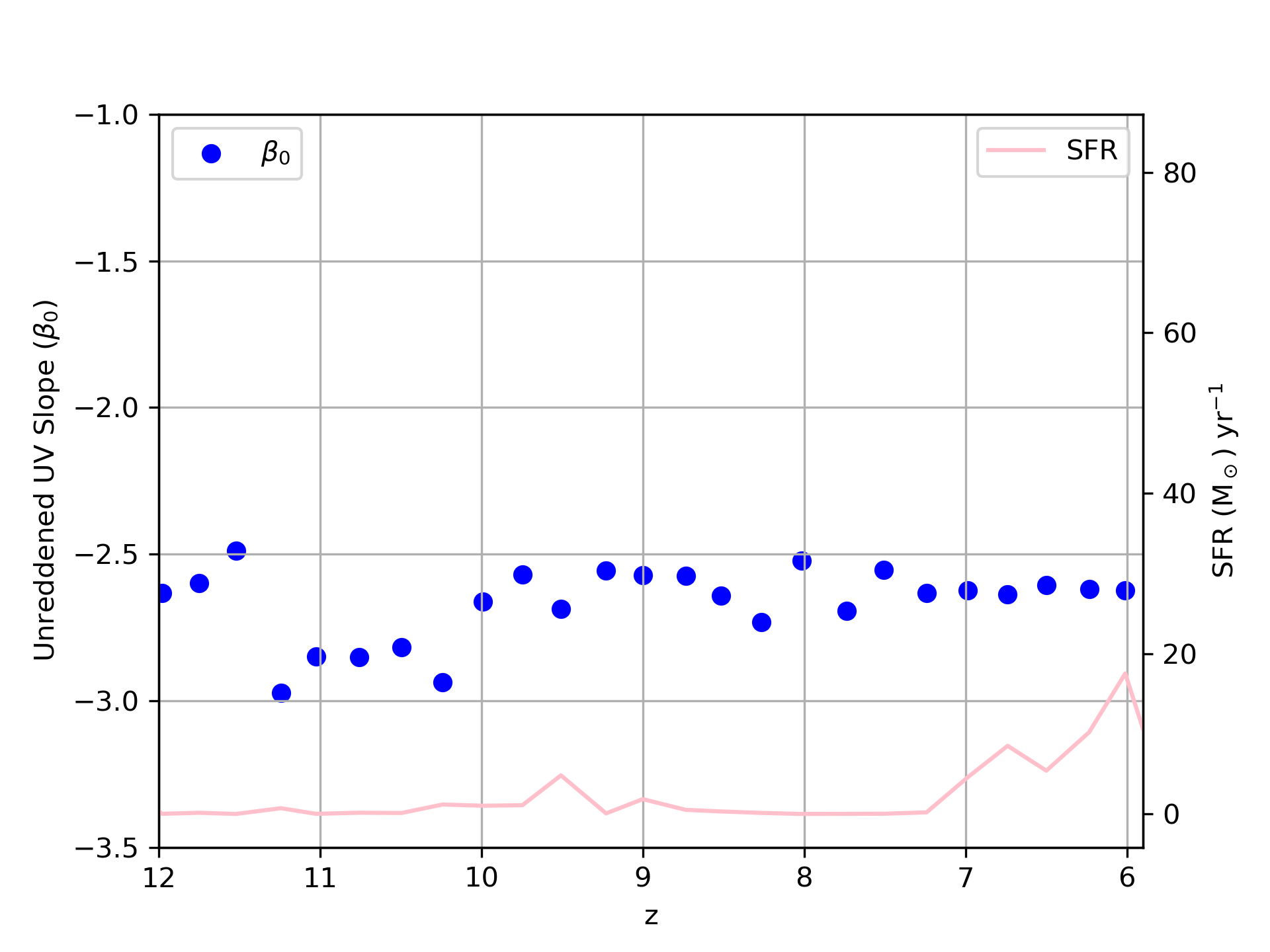}
  \caption{Unreddened UV slope ($\beta_0$) vs $z$ for galaxy h20.  See Figure~\ref{figure:beta_z}, and the associated text, for details.}
\end{figure*}

\end{document}